\documentstyle[psfig,onecolumn]{mn}
\newif\ifAMStwofonts

\newcommand{\etal}{et al. }

\newcommand{\adhoc}{{ad hoc} }
\newcommand{\adhocp}{{ad hoc}}
\newcommand{\mytorus}{{\sc mytorus} }
\newcommand{\mytorusp}{{\sc mytorus}}

\newcommand{\asca}{{\it ASCA} }

\newcommand{\rosat}{{\it ROSAT} }
\newcommand{\xmm}{{\it XMM-Newton} }

\newcommand{\chandra}{{\it Chandra} }

\newcommand{\suzaku}{{\it Suzaku} }
\newcommand{\suzakup}{{\it Suzaku}}
\newcommand{\bsax}{{\it BeppoSAX} }

\newcommand{\swift}{{\it Swift} }

\newcommand{\apec}{{\sc APEC} }

\newcommand{\olya}{O~{\sc viii}~Ly$\alpha$ }

\newcommand{\fekalfa}{{Fe~K$\alpha$} }
\newcommand{\fekalfap}{{Fe~K$\alpha$}}

\newcommand{\fekbeta}{{Fe~K$\beta$} }

\newcommand{\nika}{{Ni~K$\alpha$} }

\newcommand{\fexxv}{Fe~{\sc xxv} }
\newcommand{\fexxvp}{Fe~{\sc xxv}}

\newcommand{\thetaobs}{{$\theta_{\rm obs}$} }
\newcommand{\thetaobsp}{{$\theta_{\rm obs}$}}

\newcommand{\nhs}{{$N_{\rm H,S}$ }}
\newcommand{\nhsp}{{$N_{\rm H,S}$}}

\newcommand{\nhzp}{{$N_{\rm H,Z}$}}

\newcommand{\tablectratesp}{{Table~1}}
\newcommand{\tablehighefits}{{Table~2} }
\newcommand{\tablehighefitsp}{{Table~2}}
\newcommand{\tablemytbbfit}{{Table~3} }
\newcommand{\tablemytbbfitp}{{Table~3}}
\newcommand{\tablebbfitlines}{{Table~4} }
\newcommand{\tablebbfitlinesp}{{Table~4}}

\ifoldfss
  \ifCUPmtlplainloaded \else
    \NewTextAlphabet{textbfit} {cmbxti10} {}
    \NewTextAlphabet{textbfss} {cmssbx10} {}
    \NewMathAlphabet{mathbfit} {cmbxti10} {} 
    \NewMathAlphabet{mathbfss} {cmssbx10} {} 
  \fi
  \ifAMStwofonts
    \ifCUPmtlplainloaded \else
      \NewSymbolFont{upmath} {eurm10}
      \NewSymbolFont{AMSa} {msam10}
      \NewMathSymbol{\upi}     {0}{upmath}{19}
      \NewMathSymbol{\umu}     {0}{upmath}{16}
      \NewMathSymbol{\upartial}{0}{upmath}{40}
      \NewMathSymbol{\leqslant}{3}{AMSa}{36}
      \NewMathSymbol{\geqslant}{3}{AMSa}{3E}

    \fi
  \fi
\fi 

\ifCUPmtlplainloaded \else
  \ifAMStwofonts \else 
    \def\upi{\pi}
    \def\umu{\mu}
    \def\upartial{\partial}
  \fi
\fi

\title{A Compton-thin Solution for the \suzaku X-ray Spectrum of the Seyfert~2 Galaxy Mkn~3}
\author[T. Yaqoob \etal]{T. Yaqoob$^{1,2}$, M. M. Tatum$^{2}$, A. Scholtes$^{2}$, A. Gottlieb$^{2}$, T. J. Turner$^{2}$ \\
	$^1$Department of Physics and Astronomy, Johns Hopkins University, 3400 N. Charles St., Baltimore, MD 21218. \\
	$^2$Department of Physics, University of Maryland Baltimore County, Baltimore, MD 21250.\\
}
\date{
      Accepted 2015 August 28.  Received 2015 August 27; in original form 2015 June 15 }

\pagerange{\pageref{firstpage}--\pageref{lastpage}}
\pubyear{2015}

\begin{document}

\maketitle

\label{firstpage}

\begin{abstract} 

Mkn~3 is a Seyfert~2 galaxy that is widely regarded as an exemplary Compton-thick AGN. We study the
\suzaku X-ray spectrum using models of the X-ray reprocessor that self-consistently account
for the \fekalfa fluorescent emission line and the associated Compton-scattered, or reflection, continuum.
We find a solution in which the average global column density,
$0.234^{+0.012}_{-0.010} \times 10^{24} \ \rm cm^{-2}$, is very different
to the line-of-sight column density, $0.902^{+0.012}_{-0.013} \times 10^{24} \ \rm cm^{-2}$.
The global column density is $\sim 5$ times
smaller than that required for the matter distribution to be Compton-thick.
Our model accounts for the profiles of the \fekalfa and \fekbeta lines, and the Fe~K edge remarkably well,
with a solar abundance of Fe. The matter distribution could consist of
a clumpy medium with a line-of-sight column density higher than the global average. A uniform, spherically-symmetric
distribution alone cannot simultaneously produce the correct fluorescent line spectrum and
reflection continuum. Previous works on Mkn~3, and other AGN,
that assumed a reflection continuum from matter with an infinite column density
could therefore lead to erroneous or ``puzzling'' conclusions if the matter out of the line-of-sight is really
Compton-thin. Whereas studies of samples of AGN have generally
only probed the line-of-sight column density, with simplistic, one-dimensional models,
it is important now to establish the global column densities in AGN.
It is the global properties that affect the energy budget in terms of reprocessing of X-rays
into infrared emission, and that constrain population synthesis models of the cosmic X-ray
background.

\end{abstract}


\section{Introduction}
\label{intro}

The circumnuclear matter surrounding the putative supermassive accreting black hole in
active galactic nuclei (AGNs), thought to be responsible for obscuration of the
primary continuum emission in type~2 AGNs, also plays a critical role in accounting
for the shape of the cosmic X-ray background spectrum, or CXRB (e.g. Gilli, Comastri \& Hasinger 2007; 
Ueda \etal 2014). 
In recent years, much emphasis 
has been placed upon Compton-thick AGNs, formally defined by an optical depth to electron scattering
in the Thomson limit that is greater than unity 
($N_{\rm H} > (1.2 \sigma_{T})^{-1}$, or $1.25 \times 10^{24} \ \rm cm^{-2}$). Although this
value of $N_{\rm H}$ is arbitrary in terms of the effects of photoelectric absorption, Compton scattering,
and the prominence of the ubiquitous \fekalfa emission line, the distinction separates sources in
which the mean number of scatterings per photon is less than or greater than one (for
Compton-thin and Compton-thick sources respectively). Strictly speaking, the 
Compton-thick definition should also be energy-dependent, but the conventional use of the term in the
literature ignores this, so the conventional use implicitly refers to the low-energy limit,
when the Klein-Nishina cross-section is equal to the Thomson cross-section. With these caveats,
it is generally thought that Compton-thick AGNs could constitute $50\%$ or more of the obscured AGN
population that contributes to the CXRB (e.g. Gilli \etal 2007; Ueda \etal 2014, and references therein).

In type~1 AGNs the same circumnuclear matter distribution can imprint signatures of Compton scattering 
(reflection) and fluorescent line emission on the observed X-ray continuum. Such a scenario
is in line with AGN unification schemes, in which the line-of-sight to type~1 AGN is unobscured
due to the geometry of the circumnuclear matter (e.g. Antonucci \& Miller 1985;
Antonucci 1993; Urry \& Padovani 2005). Regardless of AGN classification, the
gas and dust can reprocess the impinging intrinsic X-ray continuum into infrared emission 
(e.g. Elitzur 2008; Yaqoob \& Murphy 2011a; Georgantopoulos \etal 2011).
Hereafter, we refer to the circumnuclear matter as the X-ray reprocessor.

Using X-ray spectroscopy to infer the properties of the X-ray reprocessor in AGNs is thus
important, yet it has not been fully exploited. This is because models used to fit the
data have traditionally been over-simplistic, employing \adhoc components that do not
self-consistently account for the \fekalfa line emission and Compton-scattered continuum.
Instead of producing the \fekalfa line and reflection continuum 
in finite column density material (such as that detected in the line-of-sight),
material out of the line-of-sight was modeled using
a slab with an infinite column density, with an inclination angle 
relative to the observer fixed at an arbitrary value. 
Such models cannot therefore be used to {\it measure} the average global column density 
out of the line-of-sight, because the starting premise is that it is Compton-thick, without
compelling justification. A good fit to the data with these models does not imply uniqueness. 

Ideally,
one would like to make no assumptions about the global column density of the reprocessor,
and use the data to constrain the allowed values, including the ``pure reflection'' spectrum case
(i.e. reflection from infinitely-thick matter) as a limiting form of a more general model. Such physically-motivated
models that self-consistently treat the Compton-reflection and \fekalfa fluorescent emission-line
for finite column density matter, and that can be directly fitted to X-ray data, have recently
been introduced and applied (e.g. Ikeda, Awaki, \& Terashima 2009; Murphy \& Yaqoob 2009; 
Brightman \& Nandra 2011; Tatum \etal 2013;
Liu \& Li 2014). As well as fitting data from new X-ray observations (e.g. 
LaMassa \etal 2014; Brightman \etal 2015), it is fruitful to re-examine archival data to investigate
to what extent previous conclusions change. 

Compton reflection 
from finite column density matter produces a rich variety of possible spectra 
(that depend on geometry and viewing angle, as well as column density), and can be very different 
to the X-ray reflection spectrum from an
infinite column density slab (e.g. see Ikeda \etal 2009; 
Murphy \& Yaqoob 2009; Brightman \& Nandra 2011; Liu \& Li 2014). In particular, the so-called ``Compton hump,''
is no longer confined to the energy range of $\sim 10-30$~keV because the peak energy
of the reflection spectrum depends on column density and geometry. In addition,
the Fe abundance, which is often inferred from comparison of the Fe K edge in the reflection
spectrum and the data, is subject to change, once the column density of the
reflector is allowed to be finite and free. Intrinsic inferred luminosities are also subject
to revision. Overall, the improved methodology allows one to probe the three-dimensional
structure of the X-ray reprocessor in a unified way, as opposed to a disjoint view
afforded by one-dimensional partial covering models, combined with reflection from an
infinite column density slab with an arbitrary orientation relative to the observer.
Indeed, the term ``Compton-thick'' now becomes even more ambiguous, as one should
specify, for a particular AGN, 
whether the term refers to the line-of-sight column density, or to the 
average column density of the global
matter distribution out of the line-of-sight, or both.

The Seyfert~2 galaxy Mkn~3, which is nearby at $z=0.013509$ (Tifft \& Cocke  1988),
has been well-studied across the electromagnetic spectrum.
It is prototypical of its class, in that the optical broad emission lines are
only seen in polarized light (Miller \& Goodrich  1990; Tran 1995). Mkn~3 has
been a target of all major X-ray astronomy missions since {\it EXOSAT},
and it is currently regarded in the literature as an archetypal Compton-thick AGN.

High spectral-resolution grating studies with \chandra (Sako \etal 2000), and \xmm
(Pounds \etal 2005; Bianchi \etal 2005) have established an extremely complex soft X-ray
spectrum (below $\sim 2$~keV), rich with emission lines due predominantly to photoionized gas covering
a broad range of ionization states. 
Hard X-ray spectra of Mkn~3 obtained with \asca (Iwasawa \etal 1994; Griffiths \etal 1998), \bsax (Cappi \etal 1999), 
\xmm (Pounds \etal 2005; Bianchi \etal 2005; Akylas, Georgantopoulos, \& Nandra 2006;
Guainazzi \etal 2012), \swift (Guainazzi \etal 2012),
and \suzaku (Awaki \etal 2008; Fukazawa \etal 2011) were
all interpreted using a similar model, consisting of line-of-sight absorption and
scattering, along with an X-ray reflection continuum formed in an
infinite column density slab of neutral material. 
The prominent \fekalfa line was modeled with an \adhoc Gaussian component.
\chandra high spectral resolution
grating data (Shu, Yaqoob, \& Wang 2011) provided the best evidence that the \fekalfa line
originates in neutral matter. Shu \etal (2011) also measured a width of 
$3140^{+870}_{-660} \ {\rm km \ s^{-1}}$~FWHM for the \fekalfa line, 
establishing the size of the line emitter to
be $\sim 4$ times larger than the optical BLR (but smaller than the 
warm, photoionized region contributing to the soft X-ray spectrum).   
The line-of-sight column density was $\sim 10^{24} \ \rm cm^{-2}$ in all of
the observations. Thus, whilst the nearly Compton-thick column density of the 
line-of-sight absorption was fairly robust because it does not depend
on the geometry of the reprocessor, the Compton-thick nature of the {\it global} 
matter distribution was an assumption, not an inference. 

The \adhoc treatment
of the \fekalfa line in previous studies also lead to some anomalous results that were deemed ``puzzling.''
For example, Guainazzi \etal (2012) found in a variability study that the 
\fekalfa line flux was uncorrelated with the X-ray reflection continuum, yet the
two emission components are physically related, implying that the model is incomplete.
Another example is the Compton shoulder to the \fekalfa line (due to scattered
line photons), whose strength relative to the line core depends on the column density
and geometry of the line emitter (e.g. see Matt 2002; Yaqoob \& Murphy 2010). 
Using a nonphysical (Gaussian) model of the Compton shoulder,
Pounds \etal (2005) forced insertion of a Compton shoulder that would be expected
from the infinite column density slab that produces the reflection continuum
in their model, at the
expense of distorting the centroid of the line core. They found that the peak of the
\fekalfa line core increased by $\sim 30$~eV to $6.43\pm0.01$~keV, and they
estimated the ratio of the flux in the Compton shoulder to that in the core
to be $\sim 20\%$. However, they also stated that adding the Compton shoulder did not
result in a statistically significant improvement to the fit. 
This lack of a compelling Compton shoulder is only puzzling if the reflection
continuum is modeled as originating in Compton-thick matter. However, such a
choice for the reflection continuum is an assumption, not an inference.
Indeed, none of the above anomalies can ever arise if self-consistent, 
finite column density models of the X-ray reprocessor are
employed, because the reflection continuum and the \fekalfa line, including its
Compton shoulder, are all self-consistently determined from the outset. 
Moreover, the Compton-thick limit can be included in such models. Ikeda \etal (2009)
did attempt to fit the \suzaku Mkn~3 spectrum with a self-consistent toroidal model.
However, their model was based on Monte Carlo calculations with insufficient statistical
quality to properly perform $\chi^{2}$ minimization and statistical error analysis, so
they were not able to derive robust constraints on the global matter distribution. 
They obtained a line-of-sight column consistent with previous studies, and were able
to fit the data with a Compton-thick global matter distribution, but they could  
not rule out smaller column densities for the matter out of the line-of-sight.

Thus, it is in the context described above that we re-examine the \suzaku X-ray spectrum of Mkn~3,
using the X-ray reprocessor models of Murphy \& Yaqoob (2009), and  Brightman \& Nandra (2011),
which overcome the limitations of previous spectral-fitting analyses. Amongst the
historical X-ray data sets, the \suzaku data are best suited for studying the X-ray reprocessor
because the instrumentation simultaneously covers the critical Fe~K band ($\sim 6-8$~keV) with
good spectral resolution and throughput (due to the XIS CCD detectors), and the hard X-ray
band above 10~keV with good sensitivity. The simultaneous broadband coverage helps to reconcile the 
absorbed and reflected continua with the \fekalfa line emission. 

The paper is organized as follows. In \S\ref{obsdata} we describe the
basic data, and reduction procedures. 
In \S\ref{spfitting} we describe the overall strategy of the analysis
that we will present, including detailed procedures for setting
up the various X-ray reprocessing models
for spectral fitting. In \S\ref{highenergyfits}
we give the results from fitting different
X-ray reprocessor models that correspond to very
distinct physical scenarios, using only data above 2.4~keV. In \S\ref{fullbandfit} we 
present the results of extending the bandpass to include the soft X-ray spectrum down to $\sim 0.5$~keV,
using the best model fitted to the high-energy data.
In \S\ref{summary} we summarize our results and conclusions.

\section{\suzaku Observation and Data Reduction}
\label{obsdata}

The present study pertains to an observation of 
Mkn~3 made by the joint Japan/US X-ray astronomy satellite, \suzaku (Mitsuda \etal~2007)
in 2005, October 22. \suzaku carries four X-ray Imaging Spectrometers (XIS -- Koyama \etal~2007) and
a collimated Hard X-ray Detector (HXD -- Takahashi \etal~2007).
Each XIS consists of four CCD detectors
at the focal plane of its own thin-foil X-ray telescope
(XRT -- Serlemitsos \etal~2007), and has a field-of-view (FOV) of $17.8' \times 17.8'$.
One of the XIS detectors (XIS1) is
back-side illuminated (BI) and the other three (XIS0, XIS2, and XIS3) are
front-side illuminated (FI). The bandpass of the FI detectors is
$\sim 0.4-12$~keV and $\sim 0.2-12$~keV for the BI detector. The useful
bandpass depends on the signal-to-noise ratio of the source since
the effective area is significantly diminished at the extreme ends of the
operational bandpasses. Although the BI CCD has higher effective area
at low energies, the background level across the entire bandpass is higher
compared to the FI CCDs. The HXD consists of
two non-imaging instruments (the PIN and GSO -- see Takahashi \etal~2007)
with a combined bandpass of $\sim 10-600$~keV.
Both of the HXD instruments are background-limited,
more so the GSO, which has a
smaller effective area than the PIN. 
For AGNs, the source
count rate is typically much less than the background and in the
present study we used only the PIN data, as the GSO data did not provide
a reliable spectrum.
In order to obtain a background-subtracted spectrum,
the background spectrum must be modeled as a function of energy and time.
The background model for the HXD/PIN has an advertised 
systematic uncertainty of 1.3\%\footnote{http://heasarc.gsfc.nasa.gov/docs/suzaku/analysis/pinbgd.html}.
However, the signal is background-dominated, and the
source count rate may be a small fraction of the background count rate,
so the net systematic error in the background-subtracted spectra could
be significant. The observation of Mkn~3 was optimized for the HXD
in terms of positioning the source at the aim point for the HXD (the so-called
``HXD-nominal pointing'') which gives a somewhat lower count-rate in the XIS
than the ``XIS-nominal'' pointing, but gives $\sim 10\%$ higher HXD effective area.

The principal data selection and screening criteria for the XIS
were the selection of only {\it ASCA} grades
0, 2, 3, 4, and 6, the removal of
flickering pixels with the
FTOOL {\tt cleansis}, and exclusion
of data taken during satellite passages through the South Atlantic Anomaly
(SAA), as well as for time intervals less than $256$~s after passages through the SAA,
using the T\_SAA\_HXD house-keeping parameter.
Data were also rejected for Earth elevation angles (ELV) less than $5^{\circ}$,
Earth day-time elevation angles (DYE\_ELV) less than $20^{\circ}$, and values
of the magnetic cut-off rigidity (COR) less than 6 ${\rm GeV}/c^{2}$.
Residual uncertainties in the XIS energy scale
are on the order of 0.2\% or less
(or $\sim 13$~eV at 6.4~keV -- see Koyama \etal~2007). 
The cleaning and data selection resulted in net exposure times
that are reported in \tablectratesp.

We extracted XIS spectra of Mkn~3 by using
a circular extraction region with a radius of 3.5'.
Background XIS spectra
were made from off-source areas of the detector, after removing
a circular region with a radius of 4.5' centered on the source,
and the calibration sources (using rectangular masks). 
The XIS spectra from all four detectors (XIS0, XIS1, XIS2, and XIS3) were combined
into a single spectrum for spectral fitting. Instrument response matrix
functions (RMF) and the ancillary response functions (ARF) were made and
combined into a single response file using standard procedures
(e.g. see appendix of Yaqoob, 2012). 
The background subtraction method for the HXD/PIN used the file 
{\tt ae100040010\_hxd\_pinbgd.evt}, corresponding to the ``tuned'' version
of the background model\footnote{See http://heasarc.gsfc.nasa.gov/docs/suzaku/analysis/pinbgd.html},

It has been known since \rosat observations of Mkn~3 that there is a
contaminating source, IXO~30, just 1.6' away (Turner, Urry, \& Mushotzky 1993;
Morse \etal 1995; Colbert \& Ptak 2002). The source has no optical counterpart
identification and its distance is unknown. It is too close to Mkn~3 to 
exclude in the \suzaku data, since the PSF has a half-power diameter that is 
comparable to the separation of Mkn~3 and IXO~30. Bianchi \etal (2005), using
\xmm data, established that IXO~30 is not variable and contributed only
7\% of the total flux in the 0.4--2~keV band (with a power-law photon index of 1.77).
Since our spectral analysis for Mkn~3 is not focussed on the soft X-ray
spectrum, this level of contamination will not have an impact on our conclusions.

The energy bandpass used for the spectrum from each instrument
was determined by the systematics of the background subtraction. 
For the XIS, the bandpass 0.53--9.3~keV was used, for which the
background was less than 50\% of the source counts in any 
spectral bin, and in the range 0.65--8.45~keV 
the background never exceeded 25\% of the source counts in any spectral bin.
The width of the bins for the XIS data was 30~eV up to 8.0~keV, and 60~eV
above 8 keV. With this spectral binning, all bins in the 0.53--9.30~keV range
had greater than 20 counts, enabling the use of the $\chi^{2}$ statistic for
spectral fitting. (We avoid grouping spectral bins using a signal-to-noise ratio
threshold because it can ``wash out'' absorption features.)
In addition to omitting data on the basis
of background-subtraction systematics, we also
omitted some spectral data that are subject to calibration uncertainties in
the effective area due to certain atomic features. It is known that
the effective area calibration is poor in the ranges $\sim 1.8-1.9$~keV 
(due to Si in the detectors) and
$\sim 2.0-2.4$~keV (Au M edges due to the telescopes). The effective
area also has a significant, steep change at $\sim 1.56$~keV 
(due to Al in the telescopes). Given that the Mkn~3   
spectrum below $\sim 2$~keV is rich with line emission 
(e.g. Sako \etal 2000; Pounds \etal 2005; Bianchi \etal 2005), 
making interpretation of modeling in the regions of calibration uncertainties difficult,
we took the conservative approach of omitting XIS data in the 1.50--2.4~keV band
for the purpose of spectral fitting. For the HXD/PIN, negative counts in the
background-subtracted spectral bins clearly indicate a breakdown of the background
model. We found that the spectral data in the energy range 14.25--45.0~keV 
produced greater than 20 counts per bin (after background subtraction),
for bin widths of 0.375~keV below
35.0~keV and 0.75~keV above 35~keV. This gives the most reliable HXD/PIN spectrum 
with minimal binning whilst at the same time qualifying the spectrum for
spectral fitting using the $\chi^{2}$ statistic.
The final energy ranges selected for spectral fitting for each instrument are
summarized in \tablectratesp, along with the corresponding count rates.

The calibration of the relative cross-normalizations of the XIS and PIN
data involves many factors, and these have been discussed
in detail in the appendix of Yaqoob (2012). For the present \suzaku observation of
Mkn~3 we have found it to be adequate to simply use the value 
of the PIN:XIS ratio, $C_{\rm PIN:XIS}$, recommended by the \suzaku Guest Observer Facility (GOF), which for
HXD-nominal observations is $C_{\rm PIN:XIS}=1.18$\footnote{ftp://legacy.gsfc.nasa.gov/suzaku/doc/xrt/suzakumemo-2008-06.pdf}.

\begin{table}
\caption[Exposure times and count rates for the Mkn~3 Suzaku spectra]
{Exposure times and count rates for the \suzaku spectra}
\begin{center}
\begin{tabular}{lcccc}
\hline

Detector & Exposure & Energy Range & Rate$^{a}$ & Percentage of \\
         &  (ks)   & (keV)    &  (count $\rm s^{-1}$) & on-source rate$^{b}$ \\
\hline
& & &  \\
XIS & 87.7 & 0.5--1.5, 2.4--9.3 & $0.1739 \pm 0.0008$ & $90.4\%$  \\
PIN & 85.5 & 14.0--45.0 & $0.1462\pm0.0025$ & $29.8\%$ \\
& & &  \\
\hline
\end{tabular}
\end{center}
$^{a}$ Background-subtracted count rate in the energy bands specified. For the XIS, this is the
rate per XIS unit, averaged over XIS0, XIS1, XIS2, and XIS3.
$^{b}$ The background-subtracted source count rate as a percentage of the total on-source count rate,
in the utilized energy intervals.
\end{table}

\section{Analysis Strategy and Spectral Fitting}
\label{spfitting}

\begin{figure}
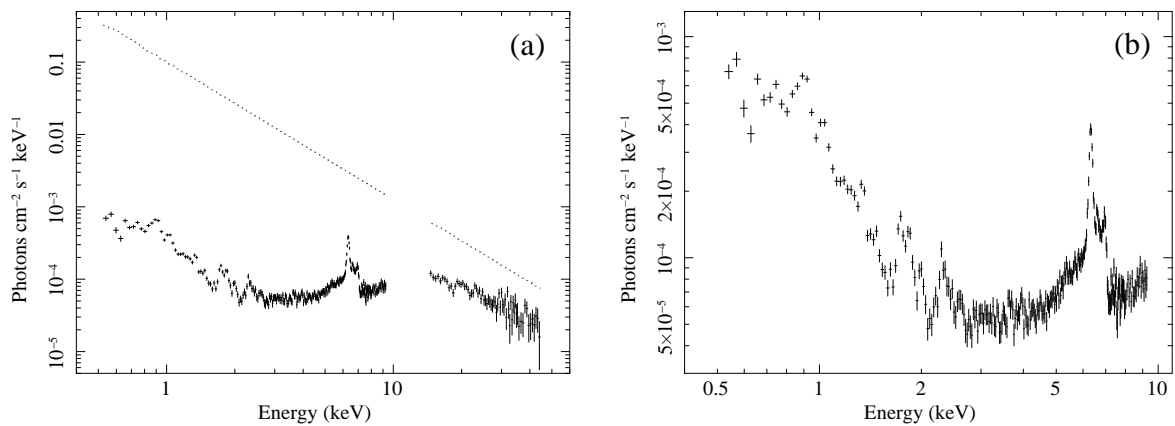

\centerline{
        \psfig{figure=f1a.ps,width=8cm,angle=270}
        \psfig{figure=f1b.ps,width=8cm,angle=270}
        }
        \caption{\footnotesize (a) The unfolded \suzaku XIS and PIN spectra of Mkn~3,
compared to a power-law with a photon index of 1.8 (dotted line). The data below and above 10 keV are
from the XIS and PIN respectively, and the data from XIS0, XIS1, XIS2, and XIS3
are combined. It can be seen that absorption and X-ray reprocessing
causes the overall spectrum above 2 keV to be very flat. (b) The same spectrum as in (a),
zoomed in on the XIS data.
}
\label{fig:plonlyufspec}
\end{figure}

In Fig.~\ref{fig:plonlyufspec} we show the unfolded \suzaku XIS and PIN spectra of Mkn~3 compared
to a simple power-law continuum with an arbitrary normalization, and photon index, $\Gamma$, of 1.8 
(a value typical of the intrinsic continuum of 
Seyfert galaxies in the pertinent energy range). The actual values of the normalization and
$\Gamma$ are not important here because the purpose of the plot is simply to show the 
salient characteristics of the overall spectrum. Full spectral-fitting will then yield the actual
parameters of the intrinsic continuum. It can be seen from Fig.~\ref{fig:plonlyufspec} that
above $\sim 2$~keV the spectrum is very flat due to absorption and reflection, and there is
complex structure around the \fekalfa emission line in the $\sim 6-8$~keV Fe~K band. The spectrum
steepens towards low energies and below $\sim 3$~keV there are many emission and absorption features,
which have been well-studied with both CCD and higher spectral resolution instrumentation 
(Sako \etal 2000; Pounds \etal 2005; Bianchi \etal 2005; Awaki \etal 2008).

The primary goal of the present spectral-fitting analysis is to determine
constraints on both the line-of-sight and global column densities of matter surrounding the X-ray
source, by modeling the \fekalfa line emission self-consistently with respect to the 
continuum characteristics associated with X-ray absorption and reflection. 
The low-energy data are generally not sensitive to column densities that are large
enough to produce the observed \fekalfa line emission, because the absorbed continuum is
suppressed at low energies and the data there are dominated by other emission components. 
Therefore we take the approach of first performing preliminary
spectral modeling of the data only above 2.4~keV, which avoids the complex spectral
features at soft X-ray energies, and also avoids the poor calibration of the
\suzaku telescope effective area in the region of the Au M-edges ($\sim 2-2.4$~keV).
After exploring different models of the X-ray reprocessor based on these high-energy fits, 
we then take the best-fitting model
and apply it to the full bandpass (excluding the $2-2.4$~keV band containing the Au~M edges),
and show that we obtain column densities that are consistent with those obtained 
from the high-energy fits that only used data above 2.4~keV. As well as the column densities,
we will show that our inferences about the constraints on the geometrical configuration of the
matter distribution around the X-ray source in Mkn~3 are also consistent between the
high-energy and full-bandpass fits.

For all of the models that we applied to the data,
we used a simple power law for the intrinsic continuum, with the
photon index ($\Gamma$) and normalization free parameters.
In the high-energy fits we do not of course need to model the extremely complex soft X-ray emission.
However, in Mkn~3, as is often
required for many other type~2 AGNs, there
is a nonthermal continuum component that is due to scattering in an extended, optically-thin
region, which carries up to a few percent of the luminosity of the intrinsic continuum.
We modeled this optically-thin scattered continuum with a power-law continuum that has the same photon index
as the intrinsic power law. We introduce a parameter $f_{s}$, which is the fraction
of the intrinsic power-law normalization that corresponds to the normalization of this scattered
component. In the optically-thin limit, $f_{s}$ is equal to the 
fraction of photons that are scattered, which in turn is approximately equal to the product of the 
scattering optical depth
and fraction of the solid angle subtended by the material at the source that is exposed
to the intrinsic X-ray continuum and visible to the observer. 
We will refer to the continuum associated with $f_{s}$ as
the ``distant scattering continuum.''

We did not apply a high-energy cutoff in the form of an exponential diminishing
factor that is often used, because such a form is unphysical. The X-ray reprocessor
models that we use instead have a termination energy for the incident continuum 
(the value depends on the model used and will be given case-by-case).
This is a closer approximation to actual X-ray spectra formed by Comptonization, which
are characterized by a power law with a single slope up to some energy, followed by a rollover, as
opposed to the continuous change in slope of exponential cutoff models.  
Moreover, the \suzaku HXD/PIN spectrum for Mkn~3 only has useful data
extending to $\sim 45$~keV (see Fig.~\ref{fig:plonlyufspec}), so
steepening of the high-energy continuum is not easily discernible, especially 
given the fact that the X-ray reprocessor models themselves cause 
their own high-energy steepening
of the observed spectra, due to Compton downscattering in the circumnuclear matter. 
From \bsax observations of Mkn~3, Cappi \etal (1999) found no evidence for a break
in the spectrum up to 150~keV. Therefore,
the use of a Comptonization model for the intrinsic continuum is not necessary. However,
it is important to note that whatever the form of the intrinsic continuum, in the model,
that continuum must extend to higher energies than the highest energy of the data,
because high-energy photons can be downscattered (many times if the medium is Compton-thick)
to within the range of the data bandpass, contributing to the observed continuum,
as well as to fluorescent line emission.

We used XSPEC (Arnaud 1996)\footnote{http://heasarc.gsfc.nasa.gov/docs/xanadu/xspec/}
for spectral fitting and utilized the $\chi^{2}$ statistic for minimization.
We included Galactic absorption with a  column density of $8.70 \times 10^{20} \ \rm cm^{-2}$
(Stark \etal 1992) in all of the model fits
described hereafter, even though it has a negligible effect for the
spectral fits above 2.4 keV. 
For all absorption components including the Galactic one, we used
photoelectric cross sections given by Verner \etal (1996).
Element abundances from Anders \& Grevesse (1989) were used throughout.
All astrophysical model parameter values will be given in the rest frame of Mkn~3
unless otherwise stated. For the sake of brevity,
certain quantities and details pertaining to particular spectral fits will be given
in the tables of results and not repeated again in the text, unless
it is necessary. Specifically, we are referring to the number of free parameters
in a fit, the number of degrees of freedom, and the null hypothesis probability.
Statistical errors given will be for one-parameter, 90\% confidence (corresponding to
a $\Delta \chi^{2}$ criterion of 2.706).

In a given energy band in the observed frame, observed fluxes will be denoted by $F_{\rm obs}$, and
observed (rest-frame) luminosities will be denoted by $L_{\rm obs}$.
These quantities are
not corrected for absorption nor Compton scattering in either
the line-of-sight material, or in the circumnuclear material. 
On the other hand, intrinsic luminosities, denoted by $L_{\rm intr}$ will
be corrected for absorption and X-ray reprocessing, and the energy band associated
with a particular value of $L_{\rm intr}$ will refer to the energy range in the
rest frame of the source.
We use a standard cosmology of $H_{0} = 70 \
\rm km \ s^{-1} \ Mpc^{-1}$, $\Lambda = 0.73$, $q_{0} = 0$ throughout the paper.

\subsection{Uniform Spherical Model}
\label{spheremodel}

Here we summarize the parameters of the uniform, spherical
(fully covering) X-ray reprocessor model of Brightman and Nandra (2011; hereafter BN11), 
implemented using the XSPEC 
emission table (``atable'') {\tt sphere0708.fits} (see BN11 for details of this
table). The table was explicitly calculated by BN11 for an incident power-law
continuum, with a termination energy of 500~keV.
The BN11 spherical model is characterized by the power-law continuum photon index ($\Gamma$)
and normalization, a radial column density,
($N_{\rm H}$), and two parameters that control the abundances of elements 
relative to solar. We call the two parameters $X_{\rm Fe}$ and $X_{\rm M}$,
where the former is the Fe abundance relative to the
adopted solar value, and the latter is a single abundance multiplier for
C, O, Ne, Mg, Si, S, Ar, Ca, Cr, and Ni relative to their respective solar values.
The solar abundances adopted in the BN11 spherical
model are those of Anders \& Grevesse (1989) and the photoelectric absorption cross-sections
are those of Verner \etal (1996).
In all of the applications described in the present paper, we set $X_{\rm M}=1.0$.
The BN11 spherical model self-consistently calculates the fluorescent $K\alpha$ emission
lines of all of the above-mentioned elements, as well as the \fekbeta line. 
One restriction of the BN11 spherical model is that 
the emission lines cannot be separated from continuum components, and in particular,
neither the Compton-scattered continuum nor the fluorescent line
spectrum can be varied with respect to the directly absorbed, line-of-sight 
(zeroth-order) continuum. A consequence of this is that any time delays between
variations in the direct continuum and the reprocessed emission from
the global matter distribution (i.e. the fluorescent line emission and
Compton-scattered continuum), cannot be accommodated by the model.

\subsection{The MYTORUS Model}
\label{mytorusmodel}

The toroidal X-ray reprocessor model, {\sc mytorus}, has
been described in detail in Murphy \& Yaqoob (2009) and Yaqoob \& Murphy (2010). The baseline geometry
consists of a torus with a circular cross section, whose diameter
is characterized by the equatorial column density, $N_{\rm H}$.
The torus is illuminated by a central, isotropic X-ray source, and
the global covering factor of the reprocessor is $0.5$, corresponding to
a solid angle subtended by the structure at the central X-ray source of
$2\pi$ (which in turn corresponds to an opening half-angle of $60^{\circ}$). 
The \mytorus model 
self-consistently calculates the \fekalfa and \fekbeta fluorescent emission-line spectrum
and the effects of absorption and Compton scattering on the X-ray
continuum and line emission.
As in the case of the BN11 spherical model, the element abundances in the \mytorus model 
are those of the solar values of Anders \& Grevesse (1989), 
and the photoelectric absorption cross-sections
are those of Verner \etal (1996). However, currently, none of the element abundances 
can be varied in the \mytorus model.

The practical implementation of the \mytorus model allows
free relative normalizations between different components of
the model in order to accommodate for differences in the actual geometry
pertinent to the source
(compared to the specific model assumptions used in the original calculations), and 
for time delays
between direct continuum, Compton-scattered continuum, and fluorescent 
line photons\footnote{See http://mytorus.com/manual/ for details}. 
The zeroth-order continuum component of the model is the direct,
line-of-sight observed continuum, and is diminished compared to the
incident X-ray continuum by absorption and Compton-scattering into
directions away from the line-of-sight. It is essentially a multiplicative factor
that is independent of the geometry, and independent of the intrinsic continuum.
This multiplicative factor is implemented with a single XSPEC table
for all applications of the model (it is {\tt mytorus\_Ezero\_v00.fits}).
The Compton-scattered continuum is implemented as an XSPEC additive
table model. We used the table {\tt mytorus\_scatteredH200\_v00.fits}, which 
corresponds to a power-law incident continuum with a termination energy
of 200~keV, and a photon index ($\Gamma$) in the range 1.4--2.6. The \fekalfa and \fekbeta emission
lines are implemented with another (single) XSPEC additive table model
that is produced from the same self-consistent Monte Carlo calculations
that were used to calculate the corresponding Compton-scattered continuum table
(we used the table {\tt mytl\_V000010nEp000H200\_v00.fits}).
The parameters for each of the three tables are the normalization of the
incident power-law continuum ($A_{\rm PL}$), $\Gamma$, $N_{\rm H}$, \thetaobsp, and the redshift ($z$).  

We denote the relative normalization between the scattered
continuum and the
direct, or zeroth-order continuum, by $A_{S}$, which has
a value of 1.0 for the assumed geometry. This value also implies that
that either the
intrinsic X-ray continuum flux is constant, or, for
a variable intrinsic X-ray continuum, that the X-ray reprocessor is
compact enough for the Compton-scattered flux to respond
to the intrinsic continuum
on timescales much less than the integration time for the spectrum.
Conversely, departures from $A_{S}=1.0$ imply departure of
the covering factor from $0.5$, or time delays between the intrinsic
and Compton-scattered continua, or both. However,
it is important to note that $A_{S}$ is {\it not} simply
related to the covering factor of the X-ray reprocessor because
the detailed shape of the Compton-scattered continuum varies with
covering factor.
Analogously to $A_{S}$, the parameter
$A_{L}$ is the relative normalization of the \fekalfa line
emission, with a value of 1.0 having a similar meaning to that for $A_{S}=1.0$.
In our analysis we will set $A_{L}=A_{S}$ (relaxing this assumption would
imply significant departure from the default model, such as non-solar Fe abundance).
In practice, $A_{S}$ and $A_{L}$ are each implemented in XSPEC by a ``constant''
model component that multiplies the Compton-scattered continuum and fluorescent line
table respectively.

The \mytorus model can be applied in a number of ways that have been detailed in 
Yaqoob (2012), and LaMassa \etal (2014). 
In the simplest mode of application of the \mytorus model,
referred to as ``coupled mode'' (regardless of the form of the intrinsic continuum),
the angle made by the axis of the torus with the observer's line-of-sight
(\thetaobsp) is coupled to the column density that is intercepted
by the zeroth-order continuum. In other words, the effective geometry of
the X-ray reprocessor is {\it precisely} that assumed in the original
Monte Carlo calculations (Murphy \& Yaqoob 2009), and all of the corresponding
parameter values in the three \mytorus tables are tied together. 
However, it is possible to apply the model in ways 
(described below) that mimic scenarios
that have more complex geometries than that for the default, baseline assumptions.

In reality, the geometry of the circumnuclear structure in the source, particularly at the edges,
may not be well represented by the exact geometry that is assumed for a specific model.
In Yaqoob (2012) we showed how the \mytorus can be used in a ``decoupled mode,'' to
crudely mimic different geometries, by decoupling the
zeroth-order continuum from the inclination angle (\thetaobsp).
Since the zeroth-order continuum is independent of geometry 
(being purely a line-of-sight
quantity), the inclination angle associated with this component becomes a dummy
parameter and it is fixed at $90^{\circ}$ so that the column density
intercepting the zeroth-order continuum
is literally equal to the value of the \mytorus equatorial model column density. (In the
coupled mode, the equatorial $N_{H}$ is not equal to the line-of-sight column density
for general values of \thetaobsp.) The inclination angle for the Compton-scattered
continuum table can be interpreted as characterizing the Compton-scattered (or reflected)
continuum in terms of the relative direction of the incident continuum source. For
example, regardless of the detailed geometry, ``back-side'' reflection, whereby the
intrinsic X-ray source lies between the observer and the reflecting material,
is well-characterized by the ``face-on'' (\thetaobs = $0^{\circ}$) Compton-scattered spectrum.
On the other hand, if the absorbing/scattering material lies inbetween the
intrinsic X-ray source  and the observer, regardless of the detailed geometry, the
observed spectrum is well-characterized by the ``edge-on'' (\thetaobs $ = 90^{\circ}$) Compton-scattered spectrum
(see Yaqoob, 2012). The reflected spectrum becomes less sensitive to \thetaobs as
the column density of the reflecting matter decreases, and in the optically-thin
limit the spectra for \thetaobs $=0^{\circ}$ and \thetaobs = $90^{\circ}$
are identical. Typically, for the purpose of fitting data, the simplest scenarios
should be tried first by fixing \thetaobs at $0^{\circ}$ and then at $90^{\circ}$,
to determine whether the spectrum can be fitted with a dominant reflection continuum
corresponding to these extremes, or a combination of them.
As an example, \thetaobs $=0^{\circ}$ might correspond to a spectrum
reflected from the ``far-side'' of a patchy
matter distribution that is observed through ``holes'' in the observer's near-side
of the distribution. (See, for example, Liu \& Li (2014), who
describe Monte Carlo simulation results
for a patchy/clumpy X-ray reprocessor: a spectral-fitting model is not publicly available, however.) 
On the other hand, \thetaobs $= 90^{\circ}$ corresponds
to a scenario in which the X-ray source is embedded in a densely
populated matter distribution that has no clear line-of-sight to back-side reflection
surfaces. 

In decoupled mode, the column densities for the zeroth-order continuum and for the
reflected and fluorescent line spectra may or may not be coupled to each other. If
they are decoupled, they will be referred to as \nhzp, and \nhsp, which represent
the line-of-sight column density and the global column density (in some average sense)
respectively. Note that $A_{S}$ should not be interpreted as a covering factor.
Moreover, in a scenario in which the Compton-scattered continuum and fluorescent line flux that
is observed is dominated by back-side reflection from the inner far side of
the X-ray reprocessor through unobscured patches,
the global covering factor cannot be constrained even in principle. 
The amount of ``leakage'' due to these patches is not related to the bulk global covering factor. In fact, the
parameter $A_{S}$ is more closely related (in this application) to the
fraction of the total solid angle subtended by the X-ray reprocessor that is
punctured by ``holes.'' It should also be remembered
that $A_{S}$ includes the effects of any time delays between
the intrinsic continuum and the response of the reflection spectra.

An alternative spectral-fitting model with a different toroidal geometry to that of the
\mytorus model, due to Brightman \& Nandra (2011), is also available (see BN11). However, a detailed
study by Liu \& Li (2015) has shown that the BN11 toroidal model suffers from some erroneous calculations
of the reprocessed X-ray continuum and line spectra, so we did not apply this model. We also note that
the toroidal models of Ikeda \etal (2009) and Liu \& Li (2014) are not publicly available.

\subsection{The \fekalfa Line Energy}
\label{fekalineenergy}

In the BN11 and \mytorus models, the centroid energy of the \fekalfa line 
emission is not a free parameter
since it is explicitly modeled as originating in neutral matter.
The same is true for the \fekbeta line.
In fact, in the \mytorus model, the \fekalfa line is explicitly modeled 
as the doublet K$\alpha_{1}$ at 6.404 keV and K$\alpha_{2}$ at 6.391 keV, 
with a branching ratio of 2:1 (see Murphy and Yaqoob (2009) for details).
For Mkn~3, Shu \etal (2011) empirically measured a peak 
rest-frame \fekalfa line energy of
$6.396^{+0.007}_{-0.008}$ keV using high-spectral-resolution \chandra HEG data.
However, in practice the peaks of the \fekalfa and \fekbeta emission lines
in the \suzaku data
may be offset relative to the baseline model because of instrumental
calibration systematics and/or mild ionization. The \suzaku data are
sensitive to offsets in the \fekalfa line peak as small as $\sim 10$~eV.
Therefore, in the \mytorus model we allowed the redshift parameter 
associated with the   
\fekalfa and \fekbeta line table to vary independently of the redshift for
all the other model components (which was fixed at the 
cosmological redshift of Mkn~3). After finding the best-fitting redshift for the
line emission, the line redshift was frozen at that value before deriving
statistical errors on the free parameters of the model.
The BN11 spherical model does not allow the fluorescent lines to be separated
from the continuum, so in that case the redshift offset had to be applied to the
BN11 model continuum as well as the lines. In the tables of spectral-fitting
results that we will present, the redshift offset will be given as the effective
\fekalfa line energy offset, $E_{\rm shift}$, in the observed frame (or $(1+z)E_{\rm shift}$
in the source frame, where $z$ is the cosmological redshift). A positive shift means that
the \fekalfa line centroid energy is higher than the expected 6.400~keV.

\subsection{The \fekalfa Line Velocity Width}
\label{fekalinewidth}
Using high-spectral-resolution \chandra HEG data, Shu \etal (2011) measured the
FWHM of the \fekalfa line in Mkn~3 to be $3140^{+870}_{-660} \ {\rm km \ s^{-1}}$. However,
this is approximately half of the FWHM spectral resolution of the \suzaku XIS
detectors at the \fekalfa line energy, so the line is not resolved by \suzakup. 
Our approach with the \suzaku data is to perform the baseline 
spectral fits with the \fekalfa line width fixed at a value much less than the
XIS spectral resolution, at $100 \ {\rm km \ s^{-1}}$ FWHM (the same width is automatically
applied to the \fekbeta line). The statistical errors on the
other free parameters of the model are derived with the line width fixed at this
value, but then the line width is allowed to be free in order to derive an 
upper limit on the FWHM.

The line broadening is achieved with the {\tt gsmooth} convolution model in XSPEC, which
convolves the intrinsic line emission spectrum with a Gaussian that has a width
$\sigma(E) = \sigma_{0} (E/ 6 \ {\rm keV})^{\alpha}$, where $\sigma_{0}$ and $\alpha$ are the
two parameters of the {\tt gsmooth} model. Since the Doppler velocity width  is
$\Delta v \sim c(\sigma(E)/E)$, fixing $\alpha=1$ models a velocity width that is
independent of energy. The parameter $\sigma_{0}$ is then related to the FWHM by
FWHM~$\sim 2.354 (c/6.0) \sigma_{0} \sim 117,700\sigma_{0}$. Note that
since the BN11 spherical model does not allow the fluorescent lines to be separated
from the continuum, the {\tt gsmooth} model in this case applies the broadening
to the continuum as well as the fluorescent lines. However, the impact of the
line width on the principal parameters for the continuum components, such as
$\Gamma$ and column densities, is negligible.

\subsection{The \fekalfa Line Flux and Equivalent Width}
\label{fekalineflux}

The \fekalfa line flux is not explicitly
an adjustable parameter because the line is produced self-consistently
in both the spherical and \mytorus models of the X-ray reprocessor.
However, by isolating the emission-line table of the \mytorus model,
keeping the best-fitting model parameters, we can measure the flux of
the \fekalfa line using an energy range that excludes the \fekbeta line.
The rest-frame flux is obtained by multiplying the observed flux by ($1+z$).
The equivalent width (EW) of the \fekalfa line was calculated using the
line flux and the measured (total) 
monochromatic continuum model flux at the observed line peak energy.
The EW in the source frame was then obtained by multiplying the observed
EW by $(1+z)$. Note that the \fekalfa line flux and EW include 
both the zeroth-order and the Compton shoulder
components of the \fekalfa line.  
The \fekalfa line is not separable from the continuum in the BN11 spherical model
so we did not explicitly derive the \fekalfa line flux and EW for this model.
Note that for both the BN11 spherical model and the \mytorus model,
the \fekbeta line flux and EW are not independent of the \fekalfa line
parameters because the theoretical value of the \fekbeta to \fekalfa branching ratio is
already factored into the self-consistent Monte Carlo simulations on which
the models are based (e.g. see Murphy \& Yaqoob 2009).

Since the \fekalfa line flux and EW are not explicit parameters,
the statistical errors on them cannot be obtained
in the usual way. However, in the \mytorus fits the parameter $A_{L}$ can be temporarily
``untied'' from $A_{S}$ in order to
crudely estimate the statistical errors on the line flux
by perturbing $A_{L}$ either side of the best-fitting value.

\section{Preliminary Spectral Fits Above 2.4~keV}
\label{highenergyfits}

Below we report the results of fitting spectral models to the \suzaku data above 2.4~keV.
The primary purpose of these high-energy spectral fits is to explore the impact of different
models on the inferred circumnuclear matter distribution in and out of the line-of-sight.
More detailed spectral fitting, including additional emission lines, continuum fluxes and
luminosities, and constraints on the \fekalfa line width is deferred to
\S\ref{fullbandfit}, which describes the broadband
fit.

\subsection{Uniform Spherical Model Fits}
\label{spherefits}

In this section we give the results of applying the uniform spherical model 
of BN11 for the circumnuclear matter distribution in Mkn~3.
See \S\ref{spfitting} and \S\ref{spheremodel} for
descriptions of the model parameters.
For the sake of reproducibility, we give the exact XSPEC model expression used to
set up the model:

\begin{eqnarray}
\rm
BN11 sphere (1) \ \ model \  = \  constant<1>*phabs<2>( & & \nonumber \\ 
\rm gsmooth<3>*(atable\{sphere0708.fits\}<4>) & + & \nonumber \\
\rm constant<5>*zpowerlw<6> ) \nonumber 
\end{eqnarray}

\begin{figure}
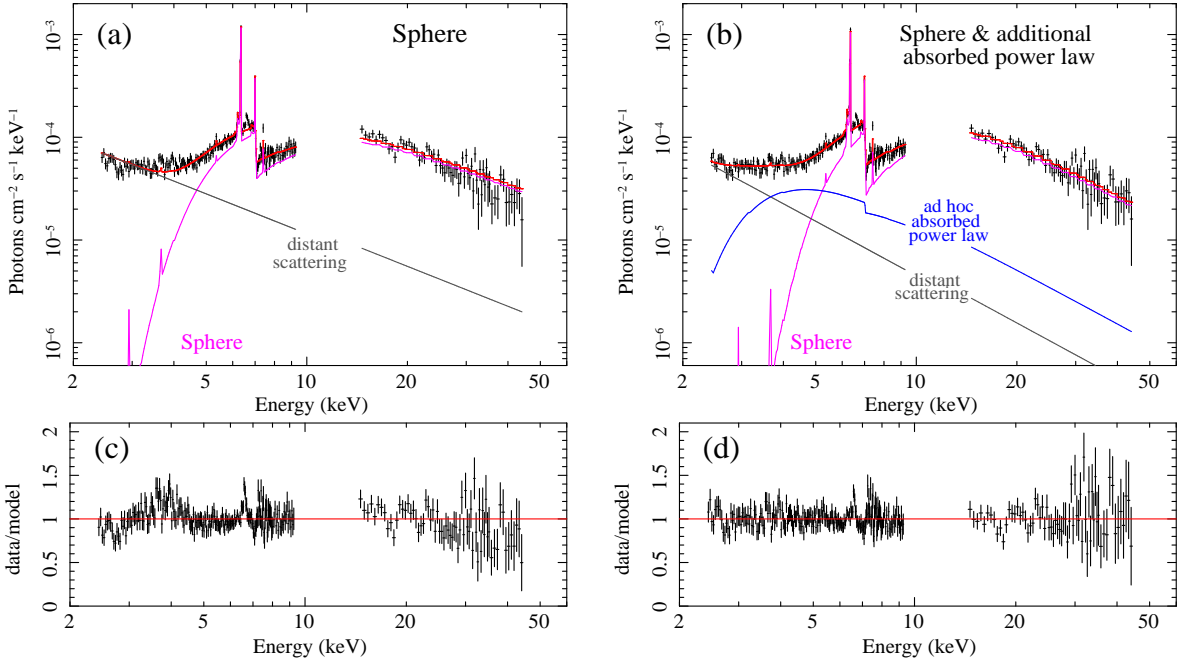

\centerline{
 \psfig{figure=f2a.ps,width=8cm,angle=270}
 \psfig{figure=f2b.ps,width=8cm,angle=270}
}
\centerline{
 \psfig{figure=f2c.ps,width=8cm,angle=270}
 \psfig{figure=f2d.ps,width=8cm,angle=270}
}
\caption{\footnotesize High-energy spectral fits to the \suzaku data for Mkn~3
with the uniform spherical X-ray reprocessor model (see \S\ref{spherefits} and \tablehighefitsp). A
continuum power-law component corresponding to optically-thin scattering in
distant matter is included in all the fits.
(a) Spherical matter distribution with no other absorption components. It can be seen that the
spectral curvature in the model and data below $\sim 5$~keV and above $\sim 25$~keV
are discrepant. (b) Spherical matter distribution with an additional (\adhocp) absorbed power-law
component. It can be seen that the discrepancies between data and model shown in (a)
are apparently resolved. (c) The data/model ratios corresponding to (a). (d) The data/model 
ratios corresponding to (b). (A color version of this figure is available in the online journal.)
}
\label{fig:highesphufspec}
\end{figure}

In the above expression, 
we identify $\rm constant<1>=C_{\rm PIN:XIS}$, $\rm phabs<2>=$ Galactic column density,
and ${\rm constant}<5> = f_{s}$ (associated with the distant scattering continuum).
There are a total of 5 free parameters and the reduced $\chi^{2}$ value for the fit is 1.732.
The results are shown in \tablehighefits (under the column ``BN11 sphere (1)''), and the 
best-fitting model overlaid on the unfolded spectrum is shown in Fig.~\ref{fig:highesphufspec}(a).
The corresponding data/model ratios are shown in Fig.~\ref{fig:highesphufspec}(c).
It can be seen that the fit is quite poor: the spectral curvature in the model below $\sim 5$~keV
does not match the data, and the power-law continuum is very flat ($\Gamma = 1.292^{+0.042}_{-0.065}$),
with the model continuum lying conspicuously above the data at energies higher than $\sim 20$~keV.
In addition, the Fe abundance relative to the solar value was free in the fit and the data forced
the relative Fe abundance to exceed the solar value ($X_{\rm Fe} = 1.30^{+0.07}_{-0.06}$).
The radial column density of the spherical matter distribution derived from the fit is 
$N_{\rm H} = 0.498^{+0.018}_{-0.021} \times 10^{24} \ \rm cm^{-2}$ (i.e. this solution is
Compton-thin).

In order to investigate whether the X-ray spectrum of Mkn~3 can be described at
all in terms of an X-ray source embedded in a fully-covering shroud of material,
we introduced an additional continuum component to the uniform spherical model
that could compensate for its deficiencies.
Namely, we added an additional power-law component that is absorbed by an additional line-of-sight
absorber. We refer to the column density of the additional absorber as $N_{\rm H,Z}$ since
it is a line-of-sight quantity. A new parameter is introduced, $f_{\rm abs}$, which is the fraction of the
intrinsic continuum that is absorbed by the additional absorber. The photon index of the
additional power-law continuum was tied to the photon index of the intrinsic continuum.
The XSPEC model expression is then:

\begin{eqnarray}
\rm
BN11 sphere (2) \ \ model \  = \  constant<1>*phabs<2>( & & \nonumber \\ 
\rm gsmooth<3>*(atable\{sphere0708.fits\}<4>) & + & \nonumber \\
\rm constant<5>*zpowerlw<6> + constant<7>*zphabs<8>*zpowerlw<9>) \nonumber 
\end{eqnarray}

where ${\rm constant<7>} = f_{\rm abs}$, and ${\rm zphabs<8>} = N_{\rm H,Z}$. 
In this fit the Fe abundance was fixed at the solar
value, so there were 6 free parameters in this model. 

The spectral-fitting results are shown in \tablehighefits (under the column ``BN11 sphere (2)''),
and the best-fitting model overlaid on the unfolded spectrum is shown in Fig.~\ref{fig:highesphufspec}(b).
The corresponding data/model ratios are shown in Fig.~\ref{fig:highesphufspec}(d).
It can be seen that the fit is much improved, showing a large reduction in the $\Delta \chi^{2}$ of 116.2.
Fig.~\ref{fig:highesphufspec}(b) and Fig.~\ref{fig:highesphufspec}(d) 
show that the continuum below $\sim 5$~keV, and above $\sim 20$~keV
is now well-fitted. The power-law index of the intrinsic continuum, and the column density of the
B11 spherical model are correspondingly larger 
($\Gamma=1.766^{+0.082}_{-0.071}$, $N_{\rm H} = 1.01^{+0.44}_{-0.12}  \times 10^{24} \ \rm cm^{-2}$).
The additional line-of-sight column density is $N_{\rm H,Z} = 0.143^{+0.028}_{-0.028} \times 10^{24} \ \rm cm^{-2}$,
and the absorbed fraction is $f_{\rm abs} = 0.0635^{+0.0081}_{-0.0071}$.
Although this modified spherical model gives an excellent fit to the data, the additional absorbed 
power-law continuum is an \adhoc component and does not have an obvious physical interpretation. The spectral
fitting results imply that the X-ray source is embedded in a thick, fully-covering matter distribution
with a column density of $\sim 10^{24} \ \rm cm^{-2}$, yet $\sim 6\%$ of the line-of-sight is less
opaque, covered by a column density that is nearly an order of magnitude smaller than the fully-covering one.
A possible interpretation of this is that the X-reprocessor is clumpy, with the clumps having
widely different column densities.

\begin{table}
\caption[Spectral-fitting results above 2.4 keV]
{Spectral-fitting results above 2.4 keV}
\begin{center}
\begin{tabular}{lcccc}
\hline
& & & & \\
Parameter & BN11 sphere (1) & B11 sphere (2) &  \mytorus & \mytorus \\
& & & (coupled) & (decoupled) \\
& & & & \\
\hline
& & & & \\
$\chi^{2}$ / degrees of freedom & 476.6/276 & 360.5/275 & 437.2/275 & 344.9/275 \\
Free Parameters &  5 & 6 & 6 & 6 \\
Reduced $\chi^{2}$ & 1.727 & 1.311 & 1.590 & 1.254 \\
Null Probability & $6.87\times 10^{-13}$ & $4.03\times 10^{-4}$ & $1.64\times 10^{-9}$ & $2.65\times 10^{-3}$ \\
$\Gamma$ & $1.292^{+0.042}_{-0.065}$ & $1.766^{+0.082}_{-0.071}$ & $1.503^{+0.111}_{-0.103}$ & $1.664^{+0.024}_{-0.071}$\\
$N_{\rm H} \rm \ (10^{24} \ cm^{-2})$  & $0.498^{+0.018}_{-0.021}$ & $1.01^{+0.44}_{-0.12}$ & $1.29^{+0.35}_{-0.35}$ & \ldots \\
$N_{\rm H, Z} \rm \ (10^{24} \ cm^{-2})$ & $\ldots$ & $0.143^{+0.028}_{-0.028}$ & \ldots & $1.000^{+0.046}_{-0.049}$ \\
Fe abundance (ratio to solar) & $1.30^{+0.07}_{-0.06}$ & 1.0(f) & 1.0(f) \\
$10^{2}f_{\rm abs}$ [sphere, additional absorbed fraction] & $\ldots$ & $6.35^{+0.81}_{-0.71}$ & \ldots & \ldots \\
$N_{\rm H, S} \rm \ (10^{24} \ cm^{-2})$ [\mytorusp] & \ldots & \ldots & \ldots & $0.221^{+0.029}_{-0.021}$ \\
\thetaobs ($^{\circ}$) & \ldots & \ldots & $64.7_{-1.7}^{+4.0}$ & \ldots \\
$E_{\rm shift}$ (eV) & $24.5^{+3.2}_{-2.8}$ & $26.8^{+3.1}_{-3.1}$ & $25.1^{+4.3}_{-2.8}$ & $21.8^{+1.7}_{-5.4}$ \\
$A_{S}$ [\mytorusp]  & \ldots & \ldots & $1.836_{-0.174}^{+0.296}$ & $0.838^{+0.041}_{-0.048}$ \\
$10^{3}f_{s}$ (optically-thin scattered fraction) & $74.7^{+7.2}_{-5.3}$ & $19.1^{+4.1}_{-3.5}$ & $40.0_{-7.4}^{+14.1}$ & $9.0^{+0.9}_{-1.0}$ \\
\hline
\end{tabular}
\end{center}
Spectral-fitting results for the \suzaku data for Mkn~3, with a uniform spherical model of the X-ray reprocessor (BN11(1)),
a uniform spherical model with an additional absorber (BN11(2)), a toroidal model with \mytorus fitted in coupled mode, and
a a toroidal model with \mytorus fitted in decoupled mode. See text for details. Fixed parameters are indicated by (f). 
Note that for the coupled \mytorus model, the lower limit on $\Gamma$ is not statistical, but corresponds to the
smallest available value of $\Gamma$ in the model tables.
The best-fitting energy shifts of the \fekalfa line model, $E_{\rm shift}$, are given at the line peak in the observed frame,
and were frozen at these values for derivation of the statistical errors on the other parameters.
\end{table}

\begin{figure}
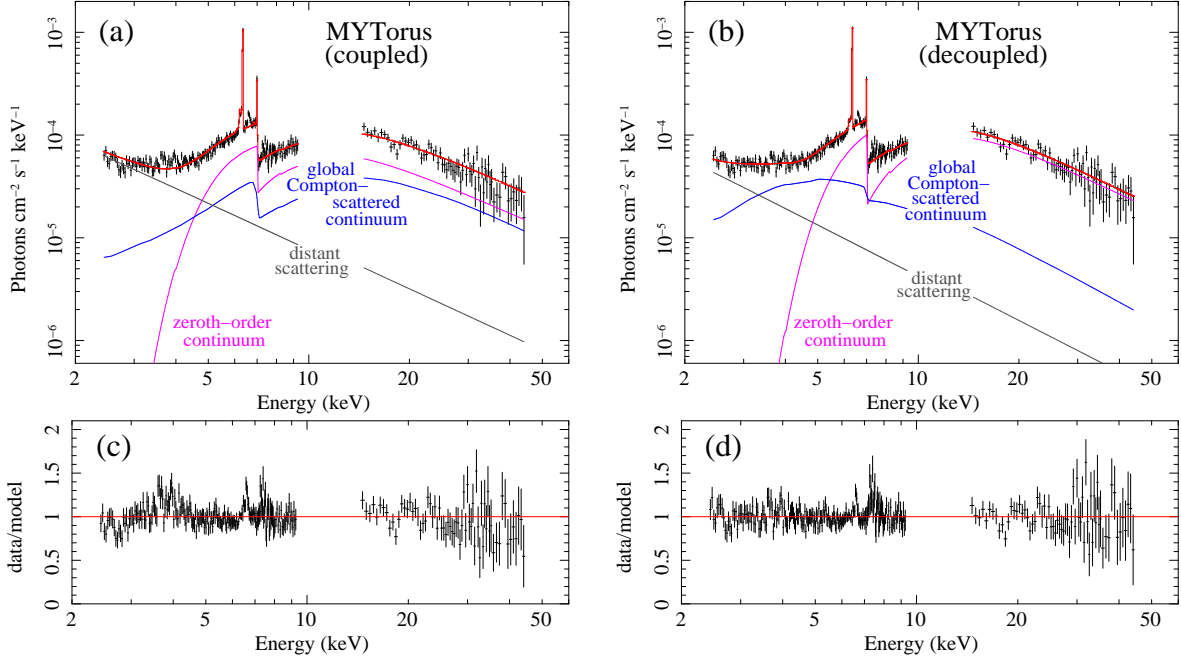

\centerline{
 \psfig{figure=f3a.ps,width=8cm,angle=270}
 \psfig{figure=f3b.ps,width=8cm,angle=270}
}
\centerline{
 \psfig{figure=f3c.ps,width=8cm,angle=270}
 \psfig{figure=f3d.ps,width=8cm,angle=270}
}
\caption{\footnotesize High-energy spectral fits to the \suzaku data for Mkn~3
with the \mytorus X-ray reprocessor model (see \S\ref{highemytfits} and \tablehighefitsp). A
continuum power-law component corresponding to optically-thin scattering in
distant matter is included in all the fits. (a) \mytorus model fit in coupled mode, with no other absorption components. 
(b) \mytorus model fit in decoupled mode, with no other absorption components. The discrepancies
between the data and spherical model shown in \ref{fig:highesphufspec}(a) are naturally resolved in
this fit by the Compton-scattered continuum
due to the same matter distribution that self-consistently fits the \fekalfa emission line,
requiring no additional, \adhocp, absorbed continuum component. 
(c) The data/model ratios corresponding to (a). (d) The data/model ratios corresponding to (b).
(A color version of this figure is available in the online journal.)
}
\label{fig:highemytufspec}
\end{figure}
\vspace{1cm}
\subsection{\mytorus Model Fits}
\label{highemytfits}

In this section we give the results of fitting the Mkn~3 \suzaku 
high-energy data with the
\mytorus model, first in coupled mode, and then in decoupled mode. The XSPEC model expression is:

\begin{eqnarray}
\rm
\mytorus \ \ model \  = \  constant<1>*phabs<2>( & & \nonumber \\ 
\rm zpowerlw<3>*etable\{mytorus\_Ezero\_v00.fits\}<4> & + & \nonumber \\
\rm constant<5>*atable\{mytorus\_scatteredH200\_v00.fits\}<6> & + & \nonumber \\
\rm constant<7>*(gsmooth<8>(atable\{mytl\_V000010nEp000H200\_v00.fits\}<9>)) & + & \nonumber \\
\rm constant<10>*zpowerlw<11>) \nonumber 
\end{eqnarray}

The model parameters have been described in \S\ref{spfitting} and \S\ref{mytorusmodel}. Here we
identify $\rm constant<1>=C_{\rm PIN:XIS}$, $\rm phabs<2>=$ Galactic column density, ${\rm constant<5>}=A_{S}$,
${\rm constant<7>}=A_{L}$, and ${\rm constant<10>}=f_{s}$. In coupled mode, the
column densities associated with each of the three \mytorus model tables
(components 4, 6, and 9 above) are tied together, as are the inclination angles. There are 6 free parameters
in the spectral fit. The spectral-fitting results are given 
in \tablehighefitsp, and the best-fitting model overlaid on the unfolded \suzaku spectrum 
is shown in Fig.~\ref{fig:highemytufspec}(a). The corresponding data/model ratios are
shown in Fig.~\ref{fig:highemytufspec}(c). The fit is similar to the fit with the pure
BN11 spherical model (see \S\ref{spherefits}, Fig.~\ref{fig:highesphufspec}(a), and Fig.~\ref{fig:highesphufspec}(c)) in the
sense that the model does not produce the correct curvature in the spectrum below $\sim 5$~keV,
and the intrinsic power-law continuum is rather flat ($\Gamma < 1.503$; note that the lower
bound in \tablehighefits is not statistical but corresponds to the lowest value of 1.400 in the model tables). 
However, the flat continuum gives
an excess relative to the data above $\sim 20$~keV, indicating that the high-energy spectrum is
steeper. The reduced $\chi^{2}$ value of 1.590 is
comparable to that obtained from the BN11 spherical model fit. The inclination angle and column density
from the coupled \mytorus fit are $64.7^{+4.0}_{-1.7} \ ^{\circ}$ and $1.29^{+0.35}_{-0.35}  \times 10^{24} \ \rm cm^{-2}$
respectively. Note that to compare this column density with that obtained from the BN11 spherical model,
the \mytorus  $N_{\rm H}$ value should be multiplied by $\pi/4$, which gives the mean column density of the torus, taking into
account the different incident angles of all rays from the intrinsic continuum source (see Murphy \& Yaqoob 2009). 
The value of the constant $A_{S}=1.836^{+0.296}_{-0.174}$ is nearly double the value
for a time-steady intrinsic X-ray continuum illuminating a torus with a covering factor of 0.5.
Thus, it is not surprising that the fit is similar to that with the fully-covering spherical model.
In Fig.~\ref{fig:highemytufspec}(a) the separate contributions to the net spectrum
are shown due to (1) the zeroth-order continuum (purple), (2) the Compton-scattered continuum from the thick matter
out of the line-of-sight (blue), and (3) the continuum from distant scattering in the optically-thin zone (grey). It can be
seen that above $\sim 5$~keV, the zeroth-order continuum dominates the net spectrum, whereas below $\sim 5$ keV,
the scattered continua dominate.

Next, we fitted the data with the \mytorus model in decoupled mode. The model expression is the same as
that for the coupled mode, but now the column density for the zeroth-order continuum, $N_{\rm H,Z}$ 
(associated with component 4), is
independent of the global column density, $N_{\rm H,S}$ (associated with components 6 and 9), responsible for producing the Compton-scattered
continuum and the fluorescent line emission. The inclination angle of the zeroth-order continuum table becomes
a dummy parameter fixed at $90^{\circ}$ (see \S\ref{mytorusmodel}). The inclination angle for
the Compton-scattered continuum and \fekalfap/\fekbeta line tables is fixed at $0^{\circ}$, which corresponds
to a scenario in which the dominant contribution to the Compton-scattered continuum and fluorescent line emission
is from the back-side of material on the far side of the X-ray source, observed through ``holes'' in the
matter distribution on the near-side to the observer. (In \S\ref{fullbandfit} we will investigate the effects
of relaxing this assumption). The spectral-fitting results are given
in \tablehighefitsp, and the best-fitting model overlaid on the unfolded \suzaku spectrum
is shown in Fig.~\ref{fig:highemytufspec}(b). The corresponding data/model ratios are shown in
Fig.~\ref{fig:highemytufspec}(d). It can be seen that the fit is significantly better than
that with the coupled \mytorus model and it is better than both of the BN11 spherical model fits 
(described in \S\ref{spherefits}). The reduced $\chi^{2}$ value is 1.254, and the
power-law photon index is $\Gamma = 1.664^{+0.024}_{-0.071}$. The spectral shape
in the regimes that were problematic for the pure spherical model (Fig.~\ref{fig:highesphufspec}(a) and
Fig.~\ref{fig:highesphufspec}(c)),
and the coupled \mytorus model (Fig.~\ref{fig:highemytufspec}(a) and Fig.~\ref{fig:highemytufspec}(c)) 
is very well reproduced by the
decoupled \mytorus model. Fig.~\ref{fig:highemytufspec}(b) shows the separate 
continuum contributions to the net spectrum using the same color-coding scheme as in Fig.~\ref{fig:highemytufspec}(a).
From this we see that the Compton-scattered continuum in the decoupled \mytorus model has
a very similar shape to the \adhoc additional absorbed power-law component that was needed to
force the spherical model to fit the data (compare the blue curves in Fig.~\ref{fig:highesphufspec}(b) and
Fig.~\ref{fig:highemytufspec}(b)). Thus, the decoupled \mytorus model naturally produces
the continuum that is required to fit the shape of the spectrum in the $\sim 3-5$~keV band,
and that continuum is produced by Compton scattering from {\it Compton-thin}
material out of the line-of-sight. The column density of that material is 
$N_{\rm H,S} = 0.221^{+0.029}_{-0.021} \times 10^{24} \ \rm cm^{-2}$ (the value obtained
from the full-band fit described in \S\ref{fullbandfit} is statistically consistent with this). The column density in
the line-of-sight is much larger, $N_{\rm H,Z} = 1.000^{+0.046}_{-0.049} \times 10^{24} \ \rm cm^{-2}$.
Thus, in this scenario, the departure from spherical symmetry is severe, with the implication
that the global, average column density is a factor of $\sim 4$ smaller than that in 
the line-of-sight. If the circumnuclear matter is in the form of a patchy distribution of
discrete clouds, with a typical column density of $N_{\rm H,S}$, the scenario described 
by the decoupled \mytorus model implies an asymmetric distribution with a 
larger mean number of clouds in the line-of-sight. Although the parameter $A_{S}$ cannot
be interpreted literally as a covering factor, the value of $A_{S} = 0.838^{+0.041}_{-0.048}$
(\tablehighefitsp) is suggestive that a global covering factor of the order of 0.5 is not unreasonable.
However, the possibility of light travel-time delays between the direct and reflected continua
means that this is by no means a robust interpretation. 

Note that the four sets of spectral-fitting results shown in \tablehighefits give
different values of $f_{s}$, in the range $\sim 0.9$--$7.5\%$. It can be seen
from Fig.~\ref{fig:highesphufspec} and Fig.~\ref{fig:highemytufspec} that $f_{s}$,
the relative magnitude of the distant-matter power-law continuum, is controlled
by how much of the data below $\sim 4$~keV are left unmodelled by the 
Compton-scattered and zeroth-order continua from the X-ray reprocessor (sphere or torus).
(Above $\sim 4$~keV the distant-matter continuum is overwhelmed by all the other
continuum components.) For example, for the pure spherical model shown in
Fig.~\ref{fig:highesphufspec}(a), the continua below $\sim 4$~keV from the X-ray reprocessor are smallest
compared to the other three fits, and indeed, $f_{s}$ is largest for the fit shown
in Fig.~\ref{fig:highesphufspec}(a) and \tablehighefits (column 2). At the other
extreme, the decoupled \mytorus model fit has the highest flux contribution below $\sim 4$~keV
from the Compton-scattered and zeroth-order continua,
and as can be see from Fig.~\ref{fig:highemytufspec}(b) and \tablehighefits (column 5),
this fit also has the lowest value of $f_{s}$ compared to the other three fits, as expected.

\section{Full-band Spectral Fit with the MYTorus Model}
\label{fullbandfit}
 
\subsection{Model Setup}
\label{decoupledsetup}

In the previous section we presented spectral-fitting results for the Mkn~3 \suzaku data 
above $2.4$~keV and showed that the decoupled \mytorus model gave the best description of
the spectrum. Although the spherical model with the addition of an \adhoc absorber
gave nearly as good a fit, the additional absorber breaks the self-consistency of
the model and the decoupled \mytorus model provides a better means of modeling
the implied clumpy X-ray reprocessor. The \mytorus model also has the advantage that the
line-of-sight continuum and global Compton-scattered continuum components are separable,
which is especially critical if there are long-term time delays between 
variations in the direct and reflected continua.
Here we present the spectral-fitting results for the decoupled \mytorus model
applied to the broadband \suzaku data (i.e. with the XIS data extended down to $0.54$~keV).
Including the lower energy data means that additional model components are required. 
Firstly, the optically-thin scattered continuum may be subject to additional absorption
by matter that is extended on the scale of the galaxy, which is of course much larger
than the matter distribution represented by the \mytorus model. Thus, we added
an additional uniform column 
density as a free parameter, which we refer to as $N_{\rm H,1}$, 
that may have been too small to be detectable in
the spectral fits above $2.4$~keV. We also added
an optically-thin continuum emission component using the {\sc apec} model with abundances
fixed at the solar values, but with the normalization
and temperature of the thermal component ($kT_{\sc apec}$ and $A_{\sc apec}$
respectively) allowed to float in fits. This optically-thin emission component,
which is commonly observed in AGN and extended on the scale of the galaxy, is 
also {\it not} absorbed
by the material represented by the \mytorus model (or at least, we observe
only the portion that is unobscured by it). However,
we do include an additional column density as a
free parameter to allow for the possibility of absorption of the thermal component 
by material in addition to the primary
X-ray reprocessing structure (we will refer to this column density as $N_{\rm H,2}$).

The soft X-ray spectrum of Mkn~3 is extremely complex, but the spectral resolution
of the XIS is significantly worse (at all energies) than the gratings aboard \chandra and \xmm
($\sim 1-2$ orders of magnitude worse than the medium-energy \chandra grating, depending
on the energy). Previous studies of Mkn~3 using \chandra and \xmm gratings have shown numerous atomic
features, indicative of a predominantly photoionized spectrum
originating in material with a wide range in ionization states (Sako \etal 2000; Pounds \etal 2005;
Bianchi \etal 2005; Awaki \etal 2008).
Thus, when grating spectra are available, CCD data cannot contribute substantially new 
understanding of the origin of the soft X-ray spectrum since many emission and absorption
features are blended together. Nevertheless, Awaki \etal (2008) provided a detailed
analysis of the same \suzaku X-ray spectrum of Mkn~3 as discussed in the present
paper, including a comprehensive study of the atomic features found in
these data. We do not repeat that analysis here and our purpose for fitting the
XIS \suzaku soft X-ray spectrum of Mkn~3 is simply to provide an empirical parameterization 
of the data in order
to investigate whether fitting the broadband spectrum significantly
affects our conclusions about the line-of-sight and global column densities that were obtained
by fitting the high-energy data with the decoupled \mytorus model (see \S\ref{highenergyfits}). 
We found that we could obtain a good parameterization of the broadband \suzaku spectrum
by adding 11 Gaussian components to the model. Each Gaussian component is characterized by
three parameters, namely the centroid energy, integrated line flux, and the energy width. 
In the fits, the centroid energy and flux of all 11 Gaussian components were allowed to float,
with the flux allowed to become negative in order to accommodate absorption features, as opposed
to emission features. We found that 10 of the Gaussian components were unresolved and
the width of these was fixed at 100 $\rm km \ s^{-1}$~FWHM. The width of the remaining
 Gaussian component, with centroid energy $\sim 1.12$~keV, was allowed to float. 
We note that 2 of the 11 Gaussian components
model emission lines are {\it not} in the soft X-ray band but were small enough in EW 
to be omitted in the preliminary high-energy fits. These are the \fexxv resonance line,
expected at $\sim 6.7$~keV, and the \nika fluorescent line, expected at $\sim 7.47$~keV. 

The model expression for this more complex fit with the decoupled \mytorus model is then

\begin{eqnarray}
\rm
\mytorus \ \ model \  = \  constant<1>*phabs<2>( & & \nonumber \\ 
\rm zpowerlw<3>*etable\{mytorus\_Ezero\_v00.fits\}<4> & + & \nonumber \\
\rm constant<5>*atable\{mytorus\_scatteredH200\_v00.fits\}<6> & + & \nonumber \\
\rm constant<7>*(gsmooth<8>(atable\{mytl\_V000010nEp000H200\_v00.fits\}<9>)) & + & \nonumber \\
\rm constant<10>*zphabs<11>(zpowerlw<12>) + zphabs<13>(apec<14>) & + &  \nonumber \\
\rm \sum_{i=1}^{i=11}{zgauss<14+i>} ). &   & \nonumber 
\end{eqnarray}

\begin{figure}
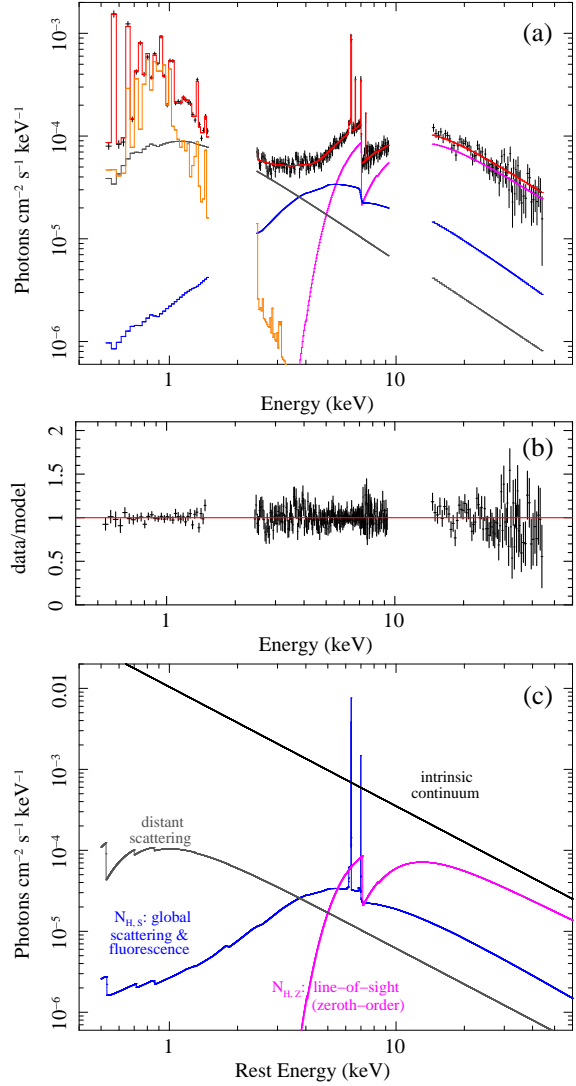

\centerline{
 \psfig{figure=f4a.ps,width=8cm,angle=270}
}
\centerline{
 \psfig{figure=f4b.ps,width=8cm,angle=270}
}
\centerline{
 \psfig{figure=f4c.ps,width=8cm,angle=270}
}
\caption{\footnotesize (a) Full, broadband spectral fit to the \suzaku Mkn~3 data with the best-fitting \mytorus model
(see \S\ref{fullbandfit}, \tablemytbbfitp, and \tablebbfitlinesp), 
showing the unfolded spectrum, the net model spectrum, and the individual model components. Note that
some emission-line features appear to be artificially narrower than they are in an unfolded spectrum.
See Fig.~\ref{fig:fullbanddatzoom} showing the same fit overlaid on the counts spectrum, which does
not suffer from this effect. (b) Data/model ratios corresponding to (a). (c) Annotated
individual continuum components of the \mytorus model fit shown in (a), the intrinsic continuum, and
the optically-thin scattered component from distant matter. Not shown in is the soft X-ray extended optically-thin
thermal emission shown in brown in (a). (A color version of this figure is available in the online journal.) 
}
\label{fig:fullbandufspec}
\end{figure}

\begin{figure}
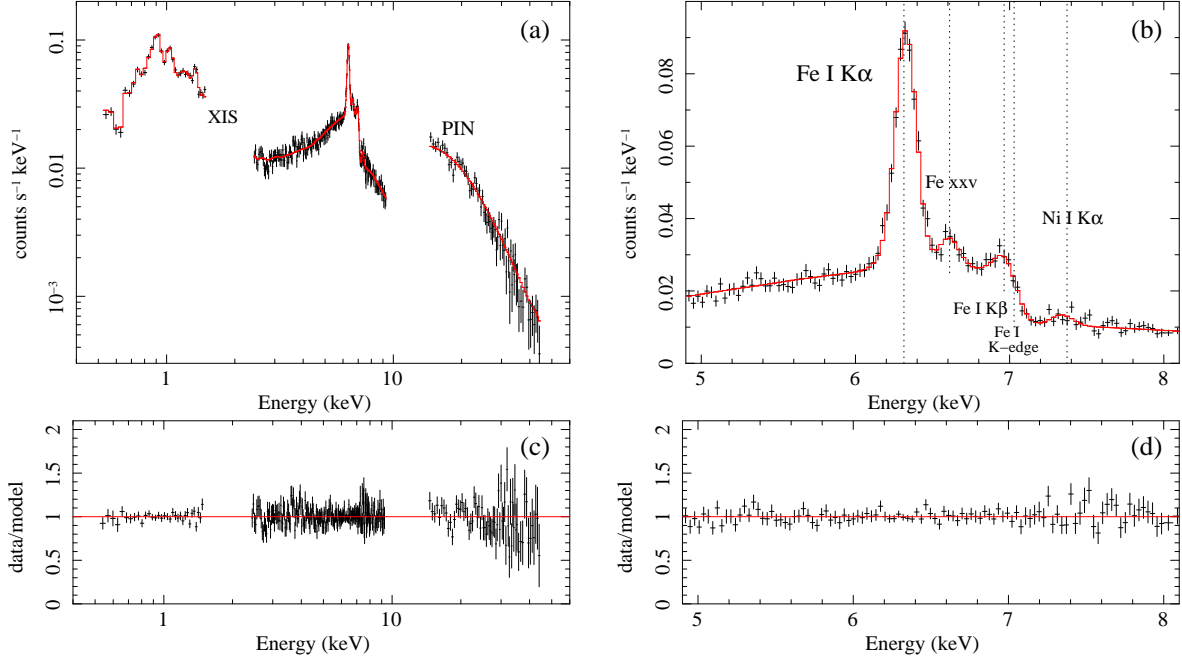

\centerline{
        \psfig{figure=f5a.ps,width=8cm,angle=270}
        \psfig{figure=f5b.ps,width=8cm,angle=270}
        }
\centerline{
        \psfig{figure=f5c.ps,width=8cm,angle=270}
        \psfig{figure=f5d.ps,width=8cm,angle=270}
        }
        \caption{\footnotesize Same broadband spectral fit as in Fig.~\ref{fig:fullbandufspec}
(see \S\ref{fullbandfit}, \tablemytbbfitp, and \tablebbfitlinesp), this time
showing the decoupled \mytorus model overlaid on the \suzaku counts spectra for Mkn~3. (a) The full broadband XIS and 
PIN data (black) overlaid with the best-fitting \mytorus model (red). Note that the energy range in the
1.5--2.4~keV band is omitted due to calibration uncertainties (see text for details). (b) The data and
model shown in (a), zoomed in on the Fe~K region, showing the detailed fit to the \fekalfa emission line
and other atomic features as labeled. Note that the flux of the model \fekalfa line is not controlled by
an arbitrary parameter but calculated self-consistently from the same matter distribution that
produces the global Compton-scattered continuum (see Fig.~\ref{fig:fullbandufspec}). The dotted lines
correspond to the expected energies of the labeled atomic features in the observed frame. The data to model
ratios corresponding to (a) and (b) are shown in (c) and (d) respectively.
(A color version of this figure is available in the online journal.)
}
\label{fig:fullbanddatzoom}
\end{figure}

There are a total of 33 free parameters, but 23 of these are associated with the Gaussian
model components, and of the remaining 10, 4 parameters are determined only by the soft X-ray
spectrum. Therefore only 6 free parameters affect the \mytorus model of the X-ray reprocessor,
the same number as in the high-energy fits described in \S\ref{highenergyfits}. The 
best-fitting values of the model parameters and their statistical errors are shown in 
\tablemytbbfit and \tablebbfitlinesp, where the latter table shows the results for
the 11 Gaussian components, and the former table shows the results for the remaining
model components. Fig.~\ref{fig:fullbandufspec}(a) shows the unfolded spectral data and model,
with the corresponding data/model ratios shown in Fig.~\ref{fig:fullbandufspec}(b).
Fig.~\ref{fig:fullbandufspec}(c) illustrates 
(in isolation of the data for greater clarity), the different continuum components that
contribute to the best-fitting model. Fig.~\ref{fig:fullbanddatzoom}(a) further illustrates
the broadband fit, showing the 
model overlaid on the counts spectrum, whilst Fig.~\ref{fig:fullbanddatzoom}(c) shows the
corresponding data/model ratios. Fig.~\ref{fig:fullbanddatzoom}(b) shows a zoom on the Fe~K band
of the model overlaid on the counts spectrum, and Fig.~\ref{fig:fullbanddatzoom}(d) shows
the corresponding data/model ratios. The reduced $\chi^{2}$ value is 1.069, and it can be seen
from Fig.~\ref{fig:fullbanddatzoom} that the fit is excellent. 

\footnotesize
\begin{table}
\caption[Results of fitting Suzaku broadband data for Mkn 3 with the mytorus model]
{Results of fitting \suzaku data for Mkn~3 with the \mytorus model}
\begin{center}
\begin{tabular}{ll}
\hline
 &  \\
Parameter &  Value \\
 & \\
\hline
$\chi^{2}$ &  300.3 \\
degrees of freedom & 281 \\
Free parameters (Gaussian components) & 23 \\
Free parameters (\mytorusp, other continua) & 10 \\
Reduced $\chi^{2}$ & 1.069 \\
Null probability & $0.206$ \\
$\Gamma$ & $1.473^{+0.007}_{-0.009}$ \\
$N_{\rm H,Z}$ ($10^{24} \ {\rm cm^{-2}}$) & $0.902^{+0.012}_{-0.013}$ \\
$N_{\rm H,S}$ ($10^{24} \ {\rm cm^{-2}}$) & $0.234^{+0.012}_{-0.010}$ \\
$A_{S}$ & $1.128^{+0.041}_{-0.030}$ \\
$10^{3} f_{s}$ (optically-thin scattered fraction) & $17.6^{+0.8}_{-0.7}$ \\
$N_{\rm H,1}$ ($10^{22} \ {\rm cm^{-2}}$) & $0.221^{+0.039}_{-0.022}$ \\
$kT_{\sc apec}$(keV) & $0.751^{+0.016}_{-0.008}$ \\
$A_{\sc apec}$ ($\rm 10^{-3} \ photons \ cm^{-2} \ s^{-1} \ keV^{-1}$) & $1.856^{+0.035}_{-0.053}$ \\
$N_{\rm H,2}$ ($10^{22} \ {\rm cm^{-2}}$) & $0.060^{+0.012}_{-0.008}$ \\
$E_{\rm shift}$ (eV) & $18.7^{+3.0}_{-3.1}$ \\
$I_{\rm Fe~K\alpha}$ ($\rm 10^{-5} \ photons \ cm^{-2} \ s^{-1}$) & $5.07^{+0.20}_{-0.18}$ \\
EW$_{\rm Fe~K\alpha}$ (eV) & $470^{+18}_{-17}$ \\
\fekalfa FWHM ($\rm km \ s^{-1}$) & 100 (f) ($<3060$) \\
$F_{\rm obs}$[0.5--2 keV] ($10^{-12} \rm \ erg \ cm^{-2} \ s^{-1}$) & $0.64$ \\
$F_{\rm obs}$[2--10  keV] ($10^{-12} \rm \ erg \ cm^{-2} \ s^{-1}$) & $6.64$ \\
$F_{\rm obs}$[10--30  keV] ($10^{-12} \rm \ erg \ cm^{-2} \ s^{-1}$) & $39.8$ \\
$L_{\rm obs}$[0.5--2 keV] ($10^{43} \rm \ erg \ s^{-1}$) & $0.26$ \\
$L_{\rm obs}$[2--10 keV] ($10^{43} \rm \ erg \ s^{-1}$) &  $2.64$ \\
$L_{\rm obs}$[10--30 keV] ($10^{43} \rm \ erg \ s^{-1}$) & $16.0$ \\
$L_{\rm intr}$[0.5--2 keV] ($10^{43} \rm \ erg \ s^{-1}$) & $2.00$ \\
$L_{\rm intr}$[2--10 keV] ($10^{43} \rm \ erg \ s^{-1}$) &  $2.47$  \\
$L_{\rm intr}$[10--30 keV] ($10^{43} \rm \ erg \ s^{-1}$) & $17.8$ \\
$L_{\sc apec}$[0.2--10 keV] ($10^{41} \rm \ erg \ s^{-1}$) & $2.0$ \\

\hline
\end{tabular}
\end{center}
Spectral-fitting results for Mkn~3 including low-energy data, described in \S\ref{fullbandfit}. The 
results for the Gaussian line components for the same fit are shown in \tablebbfitlinesp. 
The best-fitting energy shift of the \fekalfa line model, $E_{\rm shift}$, 
is given at the line peak in the observed frame,
and was frozen at that value for derivation of the statistical errors for other model parameters. The \fekalfa line
is unresolved and its width was fixed at $100 \rm \ km \ s^{-1}$~FWHM, but allowed to float for
deriving the upper limit. The observed fluxes, $F_{\rm obs}$, are total fluxes in the stated energy
bands in the observed frame. The observed and intrinsic luminosities, $L_{\rm obs}$ and $L_{\rm intr}$ 
respectively, are total luminosities in the
stated energy band in the source rest frame. All other model parameters are in the source
rest frame. The intrinsic luminosities are those for the 
incident continuum, with all absorption and reflection turned off. Some model extrapolation was employed
for fluxes and luminosities for energy intervals with no data coverage.

\end{table}
\normalsize

\begin{table}
\caption[Additional Gaussians Needed]
{Additional Gaussian components included in the \mytorus model fit to the \suzaku data}
\begin{center}
\begin{tabular}{lccccl}
\hline
& & & & \\
Line & Energy  & Flux & EW & $\Delta \chi^{2}$ & Likely ID \\
& (keV) & ($10^{-5} \rm \ photons \ cm^{-2} \ s^{-1}$) & (eV)  & \\
& & & & \\
\hline
1 & $0.580^{+0.004}_{-0.004}$ & $8.54^{+0.78}_{-1.31}$ & $949^{+87}_{-146}$ & 184.0 & O~{\sc vii} He-like triplet \\
2 & $0.674^{+0.005}_{-0.004}$ & $3.85^{+0.53}_{-0.40}$ & $321^{+44}_{-33}$ & 151.9 & \olya \\
3 & $0.748^{+0.007}_{-0.005}$ & $2.75^{+0.29}_{-0.46}$ & $32.7^{+3.5}_{-5.5}$ & 79.8 & O~{\sc vii} RRC \\
4 & $0.923^{+0.006}_{-0.007}$ & $1.64^{+0.24}_{-0.28}$ & $26.0^{+3.8}_{-4.4}$ & 196.1 & Ne~{\sc IX} He-line triplet \\
5 & $0.976^{+0.008}_{-0.007}$ & $-1.06^{+0.19}_{-0.20}$ & $-18.3^{+3.3}_{-3.5}$ & 116.5 & Ne~{\sc X}~Ly$\alpha$  \\
6 & $1.057^{+0.006}_{-0.006}$ & $1.14^{+0.15}_{-0.18}$ & $51.8^{+6.8}_{-8.2}$ & 132.8 & Fe~{\sc xxii}~L \\
7 & $1.118^{+0.081}_{-0.049}$ & $1.00^{+0.15}_{-0.19}$ & $45.5^{+6.8}_{-8.6}$ & 89.1 & Fe~{\sc xxiii}~L \\
8 & $1.268^{+0.027}_{-0.025}$ & $0.15^{+0.11}_{-0.10}$ & $8.3^{+6.1}_{-5.6}$ & 86.9 & Ne~{\sc X}~Ly$\beta$/Mg~K$\alpha$ \\
9 & $1.359^{+0.006}_{-0.009}$ & $0.55^{+0.12}_{-0.09}$ & $36.7^{+8.0}_{-6.0}$  & 23.0 & Mg~{\sc xi} He-like triplet \\
10 & $6.699^{+0.018}_{-0.019}$ & $0.70^{+0.13}_{-0.13}$ & $58.3^{+10.8}_{-10.8}$ & 96.7 & \fexxvp(r) \\
11 & $7.442^{+0.056}_{-0.050}$ & $0.32^{+0.12}_{-0.11}$ & $53.3^{+20.0}_{-18.3}$ & 6.1 & Ni~K$\alpha$ \\
\hline
\end{tabular}
\end{center}
The best-fitting parameters and statistical errors of the Gaussian model components that are part
of the model results shown in \tablemytbbfit and described in \S\ref{fullbandfit}. A negative flux
indicates an absorption feature. All except Gaussian component number 7 are unresolved: the width of each 
unresolved component was fixed at $100 \ \rm km \ s^{-1}$~FWHM. The width for component number 7 was a 
free parameter, yielding a width of $33,290_{-6,540}^{+11,700} \ \rm km \ s^{-1}$~FWHM (likely due to a blend
of two or more features). The $\Delta\chi^{2}$ values for each line correspond to the change in $\chi^{2}$
obtained when that line was removed from the model. The likely line ID's
are shown, within the systematic uncertainties in the energy scale. See Awaki \etal (2008)
for a more detailed description of the \suzaku soft X-ray line spectrum, along with theoretical expected 
values of line centroid energies.
\end{table}

\begin{figure}
\centerline{
        \psfig{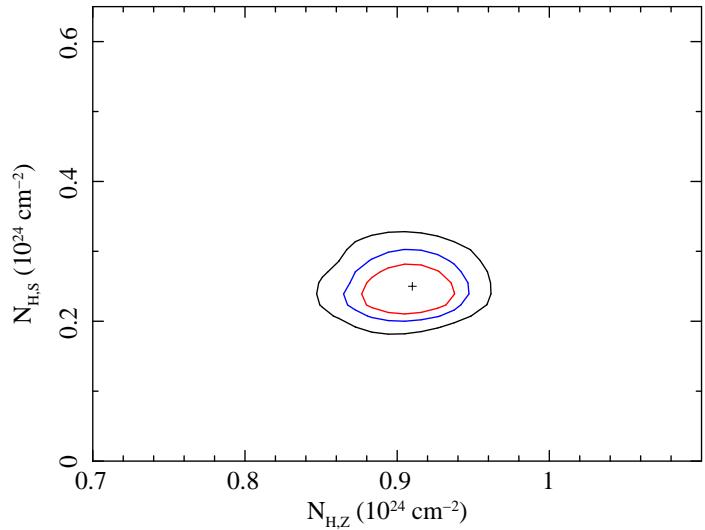}
        }
        \caption{\footnotesize Two-parameter confidence contours of the global column density, \nhsp, versus
the line-of-sight column density, \nhzp, obtained from fitting the \mytorus model to the Mkn~3 
broadband \suzaku data (see \S\ref{fullbandfit}, \tablemytbbfitp, and \tablebbfitlinesp).
The black, blue, and red contours correspond to 99\%, 90\%, and 68\% confidence respectively.
(A color version of this figure is available in the online journal.)
}
\label{fig:nhsvsnhzcontour}
\end{figure}

\subsection{X-ray Reprocessor Column Densities}

The line-of-sight column
density from the \mytorus model, \nhzp, is $0.902^{+0.012}_{-0.013} \times 10^{24} \ \rm cm^{-2}$,
and the global column density out of the line-of-sight, \nhsp, is 
$0.234^{+0.012}_{-0.010} \times 10^{24} \ \rm cm^{-2}$ (see \tablemytbbfitp). These column densities are entirely
consistent with the corresponding values obtained from the decoupled \mytorus model fitted
to the data above $2.4$~keV (\tablehighefitsp). The value of $A_{S}$ in \tablemytbbfit is
$1.128^{+0.041}_{-0.038}$, or $\sim 35\%$ higher than the value obtained from the high-energy fits.
However, it can be seen, by comparing the spectral decomposition for the two
fits in Fig.~\ref{fig:highemytufspec}(b) and Fig.~\ref{fig:fullbandufspec}(a), that
the absolute flux level of the Compton-scattered continuum is essentially the same
for the two fits. Since $A_{S}$ is a parameter that controls the flux level of the 
Compton-scattered continuum {\it relative} to the intrinsic continuum, the absolute
flux of the Compton-scattered continuum is controlled by the product of the normalization
of the intrinsic continuum and $A_{S}$. We see from \tablehighefits and \tablemytbbfit
that the photon index, $\Gamma$, of the intrinsic continuum for the broadband fit
is significantly flatter than that for the high-energy fit, and the normalization at 1~keV
of the intrinsic continuum for the broadband fit is consequently lower than that 
for the high-energy fit. Thus, a higher value of $A_{S}$ is required in the broadband
fit to maintain the same absolute flux level of the Compton-scattered continuum, and
the different values of $A_{S}$ have an insignificant impact on the inferred global column densities.
The reason for the flatter intrinsic continuum for the broadband fit will be discussed below,
in \S\ref{continua}. 

It is important to note that all of the column densities out of the line-of-sight
that we have obtained for the X-ray
reprocessor are {\it Compton-thin}, for both the BN11 spherical model
and the \mytorus model. (Recall that for the coupled \mytorus model, the equatorial
column density obtained from the fit must by multiplied by $(\pi/4)$ to obtain
the mean column density averaged over incident angles.) 
Moreover, the values of the line-of-sight column density obtained from
the various fits are either only just, or nearly Compton-thick. In the broadband fit with the
decoupled \mytorus model, the global mean column density in Mkn~3
is a long way (a factor of $\sim 5$) from being Compton-thick. Reflection from Compton-thin matter
has a different spectral shape to that from a disc with infinite column density that is
commonly used to model AGN spectra. Fig.~\ref{fig:fullbandufspec} shows that the Compton-thin
reflection component that we infer in from the \mytorus broadband fit
for Mkn~3 (blue curves) is not negligible, and in fact dominates the spectrum in the $\sim 3-5$~keV band,
it being larger than both the zeroth-order (line-of-sight) absorbed continuum and the continuum from 
distant scattering. It is principally the \fekalfa emission line and the continuum curvature below $\sim 6$~keV
that constrains this Compton-thin reflection continuum. The excellent quality of the fit in 
the Fe~K band shown in Fig.~\ref{fig:fullbanddatzoom}(c) and 
Fig.~\ref{fig:fullbanddatzoom}(d) is particularly noteworthy, considering that the \fekalfa line,
\fekbeta line, and Fe~K edge are all determined self-consistently relative to the reflection continuum
so they were not allowed to change independently of the reflection continuum. (Note that the parameter $A_{S}$
varies the reflection continuum and \fekalfa and \fekbeta lines {\it all together} and does not
allow the continuum and lines to be independent.) 
In order to further illustrate the statistical significance of the result that the line-of-sight
and global column densities in Mkn~3 obtained from the decoupled \mytorus fit 
are very different, in Fig.~\ref{fig:nhsvsnhzcontour} we show the
joint, two-parameter 68\%, 90\%, and 99\% confidence contours of \nhs versus \nhzp. It can be
seen that the constraints on both column densities are robust even at the 99\% confidence level.

In a clumpy medium the Compton-scattered reflection continuum from individual clumps
can take on a variety of shapes, depending on the average column density of the clump, 
the direction of the observer's line-of-sight relative to the direction of
the illuminating intrinsic continuum, and the obscuration by other clumps.
The shape of the reflection continuum will lie between the two extreme cases 
corresponding to the Compton-scattered photons being observed from the same
or opposite side of the clump relative to the illuminating continuum (e.g. see
Yaqoob 2012, Liu \& Lu 2014). Considering the entire clumpy matter distribution,
the first of the two extremes could correspond to a situation in which the
reflection continuum is dominated by the reflection continuum from the back-side illumination
of clumps on the opposite side of the X-ray source, relative to the observer,
reaching the observer through unblocked patches, or ``holes,'' in the near-side of the matter distribution.
The other extreme could correspond to the reflection continuum being dominated
by Compton-scattered photons escaping from the near-side of the matter distribution,
from surfaces that are on the sides opposite to the illuminated surfaces.
The latter scenario can be thought of as ``transmitted reflection.'' In the optically-thin
limit that Compton-scattered photons experience only one scattering and no
further interaction with the clumpy matter distribution, the two extreme cases 
described above will result in reflection spectra that are indistinguishable.
Thus, as the column density of the distribution is varied from being Compton-thick to
Compton-thin, the reflection continuum shape will depend less and less on the detailed
geometrical distribution, filling factor, and size of the clumps. 
Using the \mytorus model we can approximately mimic the two extremes described above
with a face-on inclination angle and an edge-on inclination angle, corresponding to 
clumps observed predominantly by
back-side reflection and ``transmitted reflection'' respectively (see Fig.~6 in Yaqoob 2012,
and the associated discussion).

In the decoupled \mytorus fit to the broadband Mkn~3 data described earlier (see \tablemytbbfitp),
the inclination angle of the torus was set to face-on, corresponding to one of the two extremes 
outlined above. From this, we found a column density that is Compton-thin, 
$N_{\rm H,S}=0.234^{+0.012}_{-0.010} \times 10^{24} \ \rm cm^{-2}$. Therefore, in light
of the earlier discussion, if we set the torus inclination angle to edge-on, we should
expect to obtain a similar column density (but note that Compton-thin does
not equate with optically-thin since the optical depth is energy-dependent).
We performed such a fit (torus inclination angle set to $90^{\circ}$) and indeed
obtained a similar column density, $N_{\rm H,S}=0.210^{+0.012}_{-0.008} \times 10^{24} \ \rm cm^{-2}$.
The fit was not as good as the face-on case but still acceptable ($\chi^{2} = 342.5$,
for 281 degrees of freedom, the same number as for the face-on case). We obtained a
line-of-sight column density of $N_{\rm H,Z}=0.997^{+0.023}_{-0.011} \times 10^{24} \ \rm cm^{-2}$,
or $\sim 10\%$ higher than the face-on case. The value of $A_{S}$ was $1.55^{+0.06}_{-0.04}$, 
or $\sim 40\%$ higher than the face-on case. The higher value is a result of
the edge-on ``transmitted reflection'' spectrum being more suppressed relative to the
intrinsic continuum than the back-side reflection spectrum for the face-on case.
We conclude that the average global column density of the X-ray reprocessor in Mkn~3 
inferred from the \mytorus fits is Compton-thin and of the order of 
$\sim 0.2-0.25 \times 10^{24} \ \rm cm^{-2}$ regardless of the detailed geometry
and orientation of the reprocessor.

\subsection{\fekalfap, \fekbeta Emission Lines, and Fe~K Edge}
\label{fekaresults}

\tablemytbbfit shows that the \fekalfa line flux and EW were measured to
be $5.07^{+0.20}_{-0.18} \ \times 10^{-5} \rm \ photons \ cm^{-2} \ s^{-1}$, and
$470^{+17}_{-18}$~eV respectively, where the EW was calculated with respect to 
the total continuum at the line peak (in the AGN frame). The \fekalfa line is unresolved with the XIS,
and the upper limit obtained on the line width (FWHM$<3060 \ {\rm km \ s^{-1}}$) 
is consistent with the \chandra HEG measurement
of $3140^{+870}_{-660} \ {\rm km \ s^{-1}}$ reported in Shu \etal (2011).
Using the simple prescription of Netzer (1990) for relating the virialized
velocity of matter orbiting a black hole, the characteristic radius 
of the line-emitting region can be written $r \sim (4/3)(c/{\rm FWHM})^{2} r_{g}$,
where $r_{g} \equiv GM/c^{2}$ is the gravitational radius. For Mkn~3, the \chandra 
HEG \fekalfa line FWHM corresponds to $r \sim 1.22 \times 10^{4}r_{g}$, or
for a black-hole mass of $4.5 \times 10^{8} M_{\odot}$ (Woo \& Urry 2002),
$r \sim 0.27$~pc. For the \suzaku upper limit measurement of the FWHM, 
which is approximately equal to the \chandra HEG best-fit value, the
preceding values of $r$ are upper limits.

The \mytorus model self-consistently calculates the Compton shoulder of both
the \fekalfa and \fekbeta lines, but the ratio of the 
flux in the Compton shoulder to that in the 
core of the \fekalfa line for the small global column density in the
best-fitting broadband model is only $\sim 9\%$ (see Yaqoob \& Murphy 2010).
The \mytorus model also self-consistently calculates the flux and profile
of the \fekbeta line. It can be seen from the model 
overlaid on the data in the Fe~K region in Fig.~\ref{fig:fullbanddatzoom}(b),
and the corresponding remarkably flat residuals in Fig.~\ref{fig:fullbanddatzoom}(d),
that the fit to the \fekalfa line, the \fekbeta line, and the Fe~K edge
is excellent, despite the fact that no empirical adjustments were applied to
either of the Fe fluorescent lines relative to the Compton-scattered continuum, 
and the Fe abundance was fixed at the solar value.
For a given set of model parameters that determine the Compton-scattered
continuum, the Fe fluorescent lines are completely determined. In the 
decoupled \mytorus model solution, the spectrum at the Fe~K edge is dominated
by the line-of-sight extinction and not the Compton-reflection continuum.

We note that previous results for the Compton shoulder in Mkn~3 were based
on fitting the shoulder with nonphysical, \adhoc models, independent of the
flux of the \fekalfa line core.
Awaki \etal (2008) claimed a detection of the Compton shoulder in the same \suzaku
data for Mkn~3 as modeled in the present paper. They used a rectangular shape for the shoulder, with a width
of 156~eV, and found a ratio of flux in the shoulder to that of the line core of
$10\pm8\%$ (one-parameter error). The peak energy of the line core increased by 6~eV when the Compton
shoulder was included. However, they did not give a statistical significance
for the detection of the shoulder, but it can be seen from the confidence
contours of line core peak energy versus Compton shoulder to line core flux ratios in Fig.~5
of Awaki \etal (2008), that the flux of the shoulder is consistent with zero at
90\% confidence for two parameters.
Pounds \etal (2005), using \xmm data for Mkn~3, 
forced a Compton shoulder in the fit, which forced the
line core centroid energy to increase above the value for neutral Fe, to $6.43\pm0.01$~keV.
Despite reporting a ratio of the Compton shoulder to line core flux of $\sim 20\%$
(with no statistical errors given), Pounds \etal (2005) stated that inclusion of the shoulder
did not result in a statistical improvement of the fit.
We note that the expected ratio of the flux in the Compton shoulder to that in the
line core from Compton-thick matter is $\sim 25-30\%$, depending on the geometry and
orientation of the line-emitting structure (e.g. see George \& Fabian 1991; Matt 2002;
Yaqoob \& Murphy 2010).
Our best description
of the \suzaku data with the \mytorus model indicates that the matter distribution
is globally Compton-thin, so the ratio of the flux in the Compton shoulder
to that in the line core is not expected to be as high as that for an origin of the \fekalfa line
in Compton-thick matter. However, even if the \fekalfa line were
formed in Compton-thick matter, it has been shown in Yaqoob \& Murphy (2010)
that if the velocity broadening of the \fekalfa line, {\it or} the
detector spectral resolution, is larger than $\sim 2000 \ {\rm km \ s^{-1}}$~FWHM,
the Compton shoulder will be rendered too smeared to be detectable, even in principle.

\subsection{Continua, Fluxes, and Luminosities}
\label{continua}

\tablemytbbfit shows that the photon index of the intrinsic power-law continuum
for the broadband decoupled \mytorus fit
is $\Gamma = 1.473^{+0.007}_{-0.009}$, which is rather flat. 
The high-energy fit with the same model gave $\Gamma =1.664^{+0.024}_{-0.071}$
(\tablehighefitsp, column 5). Amongst the high-energy
fits described in \S\ref{highenergyfits}, only the pure spherical model 
gave an intrinsic continuum flatter than this (\tablehighefitsp, column 3).
The reason for the flatter intrinsic continuum in the broadband fit is that
the steep rise in the observed spectrum below $\sim 2$~keV 
(see Fig.~\ref{fig:fullbandufspec}) is accommodated by
the thermal continuum emission (\apec model component) and this forces
the intrinsic continuum at low energies to be pushed down in flux. Without
the soft thermal component, a single, steeper power-law continuum cannot simultaneously 
accommodate the soft low-energy spectrum and the hard high-energy spectrum. 
Although statistical studies have
shown that the average intrinsic photon index is characteristically $\Gamma \sim 1.8$ for
both type~1 and type~2 AGN (e.g. Dadina 2007, 2008, and references therein), those results 
do show a range in $\Gamma$ of 1.5--2.5, with a non-negligible number of outliers even outside this
range. However, it must be borne in mind that distributions of $\Gamma$ in
previous studies were
obtained from fitting the \adhoc models invoking disc reflection with infinite column density,
and unphysical exponential cutoffs applied to the high-energy intrinsic continua.
Studies of the distribution of the intrinsic power-law continuum photon index need to be
revisited, applying the new models of finite column density X-ray reprocessors,
such as described in the present paper, to large samples of AGN. Results for
high statistical quality data for individual
sources and small samples are now beginning to emerge (e.g. see Brightman \etal 2015 
and references therein), but further work is required to establish robust parameter 
distributions.

From \tablemytbbfit it can be seen that the fraction of the intrinsic
power-law continuum required to produce the optically-thin, distant-matter scattered
continuum is $f_{s}= 0.0176^{+0.0008}_{-0.0007}$, which is
within the range generally observed in Seyfert 2 galaxies (e.g. Turner \etal~1997). 
The value of $f_{s}$ obtained from the broadband decoupled \mytorus fit is 
nearly a factor of 2 larger than the corresponding value obtained from the decoupled \mytorus 
high-energy fit ($0.0090^{+0.0009}_{-0.0010}$, see \tablehighefitsp).
However, if we compare the spectral decomposition for the 
high-energy and broadband decoupled \mytorus fits shown in 
Fig.~\ref{fig:highemytufspec}(b) and Fig.~\ref{fig:fullbandufspec}(a) respectively, 
we see that the absolute flux of the distant-matter scattered continuum is
essentially the same in both cases. Since $f_{s}$ is a parameter that controls 
the flux level of the distant-matter power-law continuum {\it relative} to the 
intrinsic continuum, the absolute
flux of the distant-matter continuum is controlled by the product of the normalization
of the intrinsic continuum and $f_{s}$. The situation is very similar to
that of the parameter $A_{S}$ being different for the broadband and high-energy fits,
and as we discussed earlier,
the photon index, $\Gamma$, of the intrinsic continuum for the broadband fit
is significantly flatter than that for the high-energy fit. Consequently, the normalization 
of the intrinsic continuum at 1~keV for the broadband fit is lower than that 
for the high-energy fit. Thus, a higher value of $f_{s}$ is required in the broadband
fit to maintain the same absolute flux level of the distant-matter continuum.
The reason for the flatter intrinsic continuum for the broadband fit has already
been discussed earlier. The column density
associated with the optically-thin, distant-matter scattered continuum is 
$N_{\rm H,1} = 0.221^{+0.039}_{-0.022} \times 10^{22} \ \rm cm^{-2}$. 

Continuum fluxes and luminosities obtained from the broadband decoupled \mytorus model fit are shown in
\tablemytbbfitp, in three energy bands: 0.5--2~keV, 2--10~keV, and 10--30~keV. For fluxes,
these energy bands are in the observed frame, but for luminosities, the lower and upper energies
in each band are quantities in the source frame, so that they appear redshifted in the 
observed frame. For regions where there were no data, the model was extrapolated.
The 0.5--2~keV, 2--10~keV, and 10--30~keV fluxes are 
$0.64$, $6.64$, and $39.8 \ \times \ 10^{-12} \rm \ erg \ cm^{-2} \ s^{-1}$ respectively. 
These values are roughly in the mid-range of historical flux levels of Mkn~3 (e.g. Iwasawa \etal 1994;
Guainazzi \etal 2012).

In \tablemytbbfitp, observed luminosities are denoted by $L_{\rm obs}$
(derived directly from the best-fitting model), and 
intrinsic luminosities are denoted by $L_{\rm intr}$ (derived
from only the best-fitting intrinsic power-law continuum). The 0.5--2~keV, 2--10~keV, and 10--30~keV
values for $L_{\rm obs}$ we obtained are $0.26$, $2.64$, and $16.0 \ \times 10^{43} \ \rm erg \ s^{-1}$
respectively. The intrinsic luminosities that we obtained in the 
0.5--2~keV, 2--10~keV, and 10--30~keV bands are $2.00$, $2.47$, and $17.8 \ \times 10^{43} \ \rm erg \ s^{-1}$
respectively. These luminosities are sensitive to the value of $\Gamma$, which complicates comparison
with the corresponding luminosities in the literature that were obtained using the
\adhoc models, which yield different values of $\Gamma$. Also, since the \adhoc models
have an adjustable parameter for the Compton-thick reflection continuum amplitude,
a model that is globally Compton-thin will not necessary yield an intrinsic luminosity
that is smaller than previous analyses that assumed Compton-thick reflection.
In fact, using the same \suzaku
data as in the present study, Awaki \etal (2008) obtained a 2--10~keV intrinsic luminosity
of $\sim 1.7 \ \times 10^{43} \ \rm erg \ s^{-1}$, which is actually lower than the value of
$\sim 2.6 \ \times 10^{43} \ \rm erg \ s^{-1}$ that we obtained from the broadband \mytorus fit.
The difference can be attributed largely to the smaller value of
$\Gamma$ that we obtained from broadband \mytorus fit. This can easily be demonstrated
by calculating the luminosity of the intrinsic power law with $\Gamma=1.8$ 
(closer to the value obtained by Awaki \etal 2008), but with same  
normalization of the intrinsic power law as in the
broadband \mytorus fit. The result is a 2--10~keV luminosity that is similar to the Awaki \etal (2008)
value.

\tablemytbbfit shows that the temperature obtained for the optically-thin
thermal continuum from the {\sc apec} model is $kT_{\sc apec} = 0.751^{+0.016}_{-0.008}$~keV, typical of Seyfert~2
galaxies (e.g. Turner \etal 1997). The column density
associated with the optically-thin thermal continuum is
$N_{\rm H,2} = 0.060^{+0.012}_{-0.008} \times 10^{22} \ \rm cm^{-2}$.
\tablemytbbfit also shows that the intrinsic luminosity of the best-fitting {\sc apec} model 
alone (in the 0.2--10~keV band) is $L_{\sc apec} = 2.0  \ \times 10^{41} \ \rm erg \ s^{-1}$.
This is less than $0.04\%$ of the bolometric luminosity of Mkn~3, which is estimated
to be $\sim 5-6 \ \times 10^{44} \ \rm erg \ s^{-1}$ by Vasudevan \etal (2010). This would
appear to align with the conclusion of Brandl \etal (2006, 2007) that Mkn~3 is AGN-dominated
as opposed to starburst-dominated, based on a study of mid-infrared flux ratios.

\subsection{Gaussian Model Components}
\label{softlineresults}

\tablebbfitlines shows the results for the 11 Gaussian model components that were
included in the broadband decoupled \mytorus fit. Shown are the centroid energies, fluxes,
equivalent widths, and a $\Delta\chi^{2}$ value for each line. The latter corresponds
to the increase in $\chi^{2}$ when the associated line is removed, and the model
re-fitted. This gives some indication of the statistical significance of each line component. In this
context, we note that a value of $\Delta\chi^{2}$ of 11.83 corresponds to a confidence level 
of $3\sigma$ for two interesting parameters. The $\Delta\chi^{2}$ values (see \tablebbfitlinesp) are
much larger than this for all of the lines except for line number 11, for which 
$\Delta\chi^{2} =6.1$, indicating a confidence level of $\sim 95\%$.

The velocity width of all except one of the Gaussian model components was fixed at
$100 \rm \ km \ s^{-1}$~FWHM, 
a value much less than the instrument resolution. For the component that the 
line width was free (line number 7), we obtained $33,290_{-6,540}^{+11,700} \ \rm km \ s^{-1}$~FWHM.
However, as stated earlier, 
due to the limited spectral resolution of CCD detectors, some of the Gaussian model components may be blends of
two or more features, or even artifacts of incorrect modeling of the soft X-ray spectrum.
Our intention is not to attempt a detailed analysis of the soft X-ray
spectrum since this has already been done with higher spectral resolution instruments
(Sako \etal 2000; Pounds \etal 2005; Bianchi \etal 2005), and it has also been done for 
these same \suzaku data by Awaki \etal (2008).

The Gaussian component number 5 in \tablebbfitlines has a negative flux and represents 
an absorption line at a rest-frame energy of $0.976^{+0.008}_{-0.007}$~keV. 
Although the energy of the line is approximately
that expected for Ne~{\sc X} resonance absorption ($1.022$~keV\footnote{http://www.pa.uky.edu/~peter/atomic}), 
we note that such an absorption feature has never 
previously been reported for Mkn~3. In fact, previous studies of Mkn~3 with \chandra grating data
(Sako \etal 2000), \xmm grating data (Pounds \etal 2005; Bianchi \etal 2005), and the
\suzaku data in the present study (Awaki \etal 2008), have not reported the
detection of any absorption lines at all, only emission lines. It is very likely that
the absorption line reported in \tablebbfitlines in the present study is an artifact of
our overly simplistic modeling. In particular the combination of the optically-thin
thermal continuum, in conjunction with other continuum components, could conspire to
produce an artificial absorption trough. We note that the model used by Awaki \etal (2008)
for the same \suzaku data as in the present study, did not include an optically-thin
thermal model component, but only Gaussian model components. Also, the observed energy
of the line, $0.976^{+0.008}_{-0.007}$~keV, compared to the expected $1.022$~keV would imply a redshift,
yet such photoionized gas in AGN is characteristically observed in outflow 
(e.g. Laha \etal 2014, and references therein).

It can be seen from \tablebbfitlines that
the \nika fluorescent line energy measured from the \suzaku data, 
namely $7.442^{+0.056}_{-0.050}$~keV, is consistent with the 
theoretical value of $7.472$~keV (e.g., Bearden 1967).
The EW of the \nika line is 
$\sim 58^{+20}_{-18}$~eV, a factor of $\sim 2$ larger
than the calculated Monte Carlo value appropriate for the 
decoupled \mytorus fit (see Yaqoob \& Murphy 2011b),
suggestive of an overabundance of Ni relative to the Anders and Grevesse (1989) solar value.

\section{Summary}
\label{summary}

We have revisited the \suzaku X-ray spectrum of the Seyfert~2 galaxy Mkn~3 to investigate
the distribution of circumnuclear matter (the X-ray reprocessor), using models that
self-consistently account for the ``neutral,'' narrow, \fekalfa emission line,
and Compton reflection of the continuum and line emission.
Our methodology is able to disentangle the column density in the line-of-sight from the
global average column density. Our conclusions are summarized below.

\begin{enumerate}

\item{We tested the data against a uniform spherical distribution using the model of 
Brightman and Nandra (2011), and obtained
a radial column density of $\sim 0.50\pm 0.02 \times 10^{24} \ \rm cm^{-2}$.
However, the fit was very poor and the model
could not produce the correct continuum shape whilst simultaneously accounting for
the \fekalfa line, even when we allowed the Fe abundance to be a free parameter. The fit
can be improved by adding additional continuum components and absorbers but the model is
then no longer that of a uniform, spherically symmetric distribution.}

\item{Applying the \mytorus model of Murphy and Yaqoob (2009) in ``coupled mode,'' yielded an equatorial
column density of $\sim 1.29\pm 0.35  \times 10^{24} \ \rm cm^{-2}$ and
an inclination angle of $64.7^{+4.0}_{-1.7} \ ^{\circ}$ for the toroidal structure, but again,
the fit was not satisfactory, due to the lack of a self-consistent solution for the continuum and \fekalfa line.}

\item{Applying the \mytorus model in ``decoupled mode,'' gave an excellent fit, and the best description
of the data, yielding a global column density, \nhsp, of $\sim 0.234^{+0.012}_{-0.010} \times 10^{24} \ \rm cm^{-2}$
that is very different to the line-of-sight column density, \nhzp, of $0.902^{+0.012}_{-0.013} \times 10^{24} \ \rm cm^{-2}$.
A physical picture consistent with this model is that of a patchy or clumpy distribution in which most
lines of sight that intercept any material typically have a column density of \nhsp, but the
line-of-sight that is relevant for us happens to have a column density that is
a factor of $\sim 4$ higher than the global average. This could be due to chance fluctuations of
material migrating in and out of our line-of-sight, or to a systematic thickening of the distribution
towards an equatorial plane that is observed edge-on or close to edge-on. In the former case,
fluctuations in the line-of-sight column density should of the order of the average global
column density or less, but in the latter case the fluctuations should be larger than the
average global column density. Detailed self-consistent modeling of the \fekalfa line and 
reflection continuum with future monitoring observations with {\it ASTRO-H} could potentially
distinguish between the two scenarios.}

\item{In the above fit with the \mytorus model, the X-ray reflection continuum originates in Compton-thin matter,
and its shape is very different to the ``standard'' Compton-reflection spectrum from an
infinite disc with infinite column density, that is commonly used to fit AGN X-ray spectra.
Our fits suggest that the reflected continuum and fluorescent line emission are observed from
the back-side of matter on the ``far-side'' of the distribution, through ``holes'' on the
side closest to the observer.}

\item{In our Compton-thin reflection model for Mkn~3, the high-energy spectrum is {\it not}
reflection dominated. Above $\sim 10$~keV, the reflection continuum is negligible and
the spectrum is dominated by the line-of-sight absorption.
Therefore the high-energy spectrum is essentially the direct, intrinsic continuum.}

\item{We applied the \mytorus model described above in two energy bands,
$\sim 2.4$--45~keV, and $\sim 0.5$--45~keV, the latter including additional soft X-ray model components.
We obtained consistent results for the global and line-of-sight column densities 
for both fits. Thus, considering AGN X-ray spectra in general, one could expect to obtain
reliable constraints on the line-of-sight and global column densities, at least
to a reasonable approximation, without having to consider the 
often considerable complexities of the soft X-ray spectrum. (However, we note that 
the relative magnitude of the
continuum from distant, optically-thin scattering {\it is} highly model-dependent.)}

\item{The \suzaku data are consistent with a solar abundance of Fe. We note that the
Fe abundance cannot trivially be deduced from the magnitude of the discontinuity 
at the Fe~K edge since the ``edge-depth'' depends on the geometry and column-dependent 
radiative transfer of continuum photons.}

\item{The \mytorus model self-consistently calculates the Compton shoulder of the
\fekalfa and \fekbeta lines, so it is already included in the model. Our best description
of the \suzaku data with the \mytorus model indicates that the matter distribution
is globally Compton-thin so the flux of the Compton shoulder is only $\sim 9\%$ of the
\fekalfa line core and that the data are consistent in detail with the model.
In previous works, attempts to reconcile the \fekalfa line profile with
a Compton-reflection continuum that was assumed to originate in an infinite disc with an
infinite column density resulted in the expectation that the ratio of the
Compton shoulder flux to that in the line core should be a factor
of $\sim 1.5-3$ higher than was observed. }

\item{The flux of the \fekbeta line in the \mytorus is self-consistently determined by
the model and is not adjustable. We obtained an excellent fit to the \fekbeta line
in the data with the decoupled \mytorus model, despite no empirical adjustments of its flux
relative to the \fekalfa line.}

\item{Our analysis also addresses the ``puzzling'' lack of correlation between the
\fekalfa line flux and the magnitude of the Compton-reflection continuum 
reported by Guainazzi \etal (2012). Their analysis insisted that the reflection
continuum is from Compton-thick matter, which forced them to conclude that the
\fekalfa line originates in a separate Compton-thin medium, whilst the \fekalfa
line from the Compton-thick matter must be suppressed. However they were not
able to offer an explanation for such suppression. Guainazzi \etal (2012) also stated
that the reflection continuum from the Compton-thin matter is unobservable.
However, they did not fit Compton-thin models of the reflection continuum and 
\fekalfa line to support that assertion. In contrast, we fitted self-consistent
models that make no presumptions about the column density of matter producing
the reflection continuum and the \fekalfa line. We found that the global
matter distribution is Compton-thin and its reflection continuum is observable.
The situation of a lack of correlation between the
\fekalfa line flux and Compton-reflection can never arise in
analyses using the self-consistent models that
we applied because the physical relationship is built in from the outset. 
We also note that an argument sometimes used to support the assumption
of a Compton-thick reflection continuum is the observation of a high infrared to X-ray continuum
ratio, and this was also invoked by Guainazzi \etal (2012) for Mkn~3.
However, Yaqoob \& Murphy (2011a) have shown that the infrared to X-ray
ratio is not uniquely determined by the column density of the reprocessor, but
has just as important a dependence on its covering factor and on the hardness of
the intrinsic X-ray continuum. They further showed that it is possible for
a Compton-thin reprocessor to produce an infrared to X-ray ratio that can
be comparable to, or actually larger, than that produced by a Compton-thick reprocessor.}

\end{enumerate}

Our analysis reveals the important role played by {\it Compton-thin} X-ray reflection from
finite-column-density matter in
constraining the average global column density of the matter distribution responsible
for producing the narrow \fekalfa emission line in AGN.
Compton-thin reflection exhibits a richer variety of spectral shapes than the 
commonly-used, disc-reflection spectrum, which assumes an infinite column density.
The X-ray reflection spectrum from finite-column-density material can peak at lower energies
than the infinite-column-density counterpart. In Mkn~3, according to the 
spectral-fitting results with the \mytorus model, the Compton-thin reflection spectrum
is the dominant continuum in the $\sim 3-5$~keV band. It is noteworthy that the 
\mytorus fit to the \fekalfa line, \fekbeta line, and the Fe~K edge region,
is so good (see Fig.~\ref{fig:fullbanddatzoom}) that there is no suggestion that any more complexity
is required, in addition to the mundane, nonrelativistic, physics of the model.
The fit in the critical Fe~K region was achieved without any free parameters that adjust 
the \fekalfa line, \fekbeta line, or the Fe~K edge with respect to the
Compton-reflection continuum.

Our results are also relevant for the study of so-called ``changing-look''
AGN, which are reported to change from being Compton-thick to Compton-thin,
or vice-versa. The phenomenon has commonly been interpreted
in terms of transits of clumpy material across the line-of-sight 
(e.g. Matt, Guainazzi \& Maiolino 2003; Risaliti \etal 2010, and references therein),
but traditional modeling has been restricted by the assumption that the global
matter distribution is Compton-thick. This global distribution contributes
an underlying reflection spectrum that is often non-varying whilst the
line-of-sight absorption spectrum varies. The possibility that the global
matter distribution is Compton-thin then leads to a richer phenomenology
that could be observable in changing-look AGN. Indeed, such a case has already
been found from a \suzakup, \xmm, and \swift observation campaign of the
type~2 AGN, NGC~454 (Marchese \etal 2012). Spectral fitting with the \mytorus
model showed that a Compton-thin reflection component (from a global matter
distribution with a column density of $\sim 3 \times 10^{23} \ \rm cm^{-2}$)
remained steady whilst the line-of-sight column density varied between
$\sim 10^{23} \ \rm cm^{-2}$ and $\sim 10^{24} \ \rm cm^{-2}$.
In the ``transiting clumps'' scenario, no ``shutting-down'' or ``powering up'' 
of the central engine is
required to explain the apparent transitions: the average Compton-thin or
Compton-thick global matter 
distribution remains steady over time, only the amount of matter in the line-of-sight
fluctuates. This picture is also consistent with the findings of Markowitz, Krumpe, \& Nikutta (2014), who
investigated the properties of transient absorption events in AGN, leading to constraints
on clumpy torus models of the circumnuclear matter distribution. 
In the transiting clump model of spectral variability, changing-look AGN
are not members of a special class of AGN. Rather, they are a loosely
defined subset of AGN in which the difference between the line-of-sight and
global column densities happens to be large enough to cause transitions
between two ``extremal'' spectral states, one that is Compton-thick and the other
 Compton-thin. Other AGN, which may constitute a majority, in which the
contrast between the line-of-sight and global column densities is
less severe, would exhibit more moderate spectral variability, insufficient
to classify them as changing-look AGN. Nevertheless, the
transiting matter interpretation of changing-look AGN cannot be universal, as
LaMassa \etal (2015) have presented a detailed multiwaveband study of a type 1 quasar to type 1.9 AGN
transition that cannot be explained by variable line-of-sight absorption, but
instead is attributed to variable ionization of the matter surrounding the
central source.

The clumpy nature of the torus that is increasingly becoming apparent from
X-ray observations has also been inferred from theoretical work on the
infrared properties of AGN (e.g. Elitzur 2008; Nenkova \etal 2008, 2010). In particular,
Nenkova \etal (2008) note that observations imply a clumpy medium because it
can produce ``isotropic infrared emission but highly anisotropic obscuration.'' 
Nevertheless, it has been argued by Feltre \etal (2012), based on a detailed
study of smooth and clumpy models, that ambiguities still exist because the
two types of model depend on very different assumptions.
Regardless of the clumpiness of the X-ray reprocessor,
it will be important to determine for a larger sample of type~2 AGN,
how common it is for the mean global column density of the X-ray reprocessor to be
very different to the line-of-sight column density. Traditionally, simplistic,
one-dimensional models have been fitted to derive column densities, with the implicit
assumption that the global column density is the same as the line-of-sight column
density. Whether or not this is the case clearly has implications for the energy
budget for reprocessing X-rays into infrared emission, and for population synthesis models
of the cosmic X-ray background.

\vspace{5mm}
\noindent
Acknowledgments \\
The authors acknowledge support for this work from NASA grants 
NNX09AD01G, NNX10AE83G, and NNX14AE62G. This research has made use of data and software provided by 
the High Energy Astrophysics Science Archive Research Center (HEASARC), 
which is a service of the Astrophysics Science Division at NASA/GSFC and the 
High Energy Astrophysics Division of the Smithsonian Astrophysical Observatory.

\bsp
\label{lastpage}


\begin{thebibliography}{}

\bibitem{} Akylas A., Georgantopoulos I., Nandra K., 2006, AN, 327, 1091

\bibitem{} Anders E., Grevesse N., 1989, Geochimica et Cosmochimica Acta 53, 197

\bibitem{} Antonucci R. R. J., 1993, ARA\&A, 31, 473

\bibitem{} Antonucci R. R. J., Miller, J. S., 1985, ApJ, 297, 621 

\bibitem{} Arnaud K. A., 1996, in Astronomical Data Analysis
Software and Systems V, ed. Jacoby, G., Barnes, J.
(Astronomical Society of the Pacific), Conference Series, Vol. 101, p. 17 

\bibitem{} Awaki, H. et al., 2008, PASJ, 60, 293

\bibitem{} Bearden J. A., 1967, Rev. Mod. Phys., 39, 78 

\bibitem{} Bianchi S., Miniutti G., Fabian A. C., Iwasawa K.,
2005, MNRAS, 360, 380

\bibitem{} Brandl B. R. et al., 2006, ApJ, 653, 1129

\bibitem{} Brandl B. R. et al., 2007, ApJ, 665, 884 

\bibitem{} Brightman M., Nandra K., 2011, MNRAS, 413, 1206 (BN11)

\bibitem{} Brightman M. et al., 2015, ApJ, 805, 41

\bibitem{} Cappi M. et al., 1999, A\&A, 344, 857

\bibitem{} Colbert E. J. M., Ptak A. F., 2002, ApJS, 143, 25

\bibitem{} Dadina M., 2007, A\&A, 461, 1209

\bibitem{} Dadina M., 2008, A\&A, 485, 417 

\bibitem{} Elitzur M., 2008, NewAR, 52, 274

\bibitem{} Feltre A., Hatziminaoglou E., Fritz J., Franceschini A., 2012, MNRAS, 426, 120

\bibitem{} Fukazawa Y., et al., 2011, ApJ, 727, 19

\bibitem{} Georgantopoulos I. et al., 2011, A\&A, 534, 23

\bibitem{} George I. M., Fabian A. C. 1991, MNRAS, 249, 352

\bibitem{} Gilli R., Comastri A., Hasinger G., 2007, A\&A, 463, 79

\bibitem{} Griffiths R. G., Warwick R. S., Georgantopoulos I., Done C., Smith D. A., MNRAS, 298, 1159

\bibitem{} Guainazzi M., La Parola V., Miniutti G., Segreto A., Longinotti A. L.
2012, A\&A, 547, 31

\bibitem{} Ikeda S., Awaki H., Terashima Y., 2009, ApJ, 692, 608

\bibitem{} Iwasawa K., Yaqoob T., Awaki H., Ogasaka Y., 1994, PASJ, 46, L167

\bibitem{} Koyama K., et al., 2007, PASJ, 59, 23

\bibitem{} Laha S., Guainazzi M., Dewangan G. C., Chakravorty, S., Kembhavi A. K. 2014, MNRAS, 441, 2613

\bibitem{} LaMassa S. M., Yaqoob T., Ptak A., Jianjun J., Heckman T. M., Gandhi P.,
Urry C. M., 2014, ApJ, 787, 61

\bibitem{} LaMassa S. M. et al., 2015, ApJ, 800, 144

\bibitem{} Liu Y., Li X., 2014, ApJ, 787, 52

\bibitem{} Liu Y., Li X., 2015, MNRAS, 448, L53

\bibitem{} Marchese E., Braito V., Della Ceca R., Caccianiga A., Severgnini, P., 2012, MNRAS, 421, 1803

\bibitem{} Markowitz A., Krumpe M., Nikutta R., 2014, MNRAS, 439, 1403 

\bibitem{} Matt G., 2002, MNRAS, 337, 147

\bibitem{} Matt G., Guainazzi M., Maiolino R., 2003, MNRAS, 342, 422

\bibitem{} Miller J. S., Goodrich R. W., 1990, ApJ, 355, 456

\bibitem{} Mitsuda K. et al., 2007, PASJ, 59, 1

\bibitem{} Morse J. A., Wilson A. S., Elvis M.,  Weaver K. A. 1995, ApJ, 439, 121

\bibitem{} Murphy K. D., Yaqoob T., 2009, MNRAS, 397, 1549 

\bibitem{} Nenkova M., Sirocky M. M., Nikutta R., Ivezi\'{c} \v{Z}., Elitzur, M., 2008, ApJ, 685, 160

\bibitem{} Nenkova M., Sirocky M. M., Nikutta R., Ivezi\'{c} \v{Z}., Elitzur, M., 2010, ApJ, 723, 1827 

\bibitem{} Netzer H. 1990, in Active Galactic Nuclei,
ed. R. D. Blandford, H. Netzer, L. Woltjer (Berlin: Springer), 137

\bibitem{} Risaliti G., Elvis M., Bianchi S., Matt G., 2010, MNRAS, 406, L20


\bibitem{} Shu X. W., Yaqoob T., Wang J. X., 2011, ApJ, 738 147

\bibitem{} Serlemitsos P. J. et al., 2007, PASJ, 59, 9

\bibitem{} Takahashi T., et al., 2007, PASJ, 59, 35

\bibitem{} Tatum M. M., Turner T. J., Miller L., Reeves J. N., 2013, ApJ, 762, 80

\bibitem{} Tifft W. G., Cocke W. J, 1988, ApJS, 67, 1

\bibitem{} Tran H. D., ApJ, 440, 565

\bibitem{} Turner T. J., George I. M., Nandra K., Mushotzky R. F. 1997, ApJ, 488, 164

\bibitem{} Turner T. J., Urry C. M., Mushotzky R. F., 1993, ApJ, 418, 653

\bibitem{} Sako M., Kahn S., Paerels F., Liedahl D. A., 2000, ApJ, 543, L115

\bibitem{} Stark A. A., Gammie C. F., Wilson R. W., Bally J., Linke R.,
Heiles C., Hurwitz M., 1992, ApJS, 79, 77

\bibitem{} Ueda Y., Akiyama M., Hasinger G., Miyaki T., Watson M., 2014, ApJ, 786, 104

\bibitem{} Urry C. M., Padovani P., 1995, PASP, 107, 803

\bibitem{} Vasudevan R. V., Fabian A. C., Gandhi P., Winter L. M., Mushotzky R. F., 2010, MNRAS, 402, 1081
 
\bibitem{} Verner D. A., Ferland G. J., Korista K. T., Yakovlev D. G., 1996, ApJ, 465, 487

\bibitem{} Woo J. H., Urry C. M. 2002, ApJ, 579, 530

\bibitem{} Yaqoob T., 2012, MNRAS, 423, 3360

\bibitem{} Yaqoob T., Murphy K. D., 2010, MNRAS, 412, 277

\bibitem{} Yaqoob T., Murphy K. D., 2011a, MNRAS, 412, 835

\bibitem{} Yaqoob T., Murphy K. D., 2011b, MNRAS, 412, 1765 


\end{thebibliography}
\end{document}